\documentclass[a4paper,fleqn]{cas-sc} 

\usepackage{xcolor}
\usepackage{hyperref}
\usepackage{amsmath}
\usepackage{cleveref}
\usepackage{amsfonts}
\usepackage{amssymb}
\usepackage{graphicx}
\usepackage{array}
\usepackage{bm}
\usepackage[modulo]{lineno}
\usepackage[numbers]{natbib}
\usepackage{nomencl}
\usepackage{makecell}
\usepackage{float}
\usepackage{multirow}
\usepackage{siunitx}
\usepackage[utf8]{inputenc}
\usepackage{placeins}
\usepackage{nicefrac}
\makenomenclature
\usepackage[center]{caption} 
\usepackage[title]{appendix}
\newif\ifshowhighlight
\showhighlightfalse 

\crefname{equation}{{Eq.}}{{Eqs.}}
\crefname{figure}{{Fig.}}{{Figs.}}
\crefname{tabular}{{Tab.}}{{Tabs.}}
\crefname{section}{{Section}}{{Sections}}
\crefname{appendix}{{Appendix}}{{Appendices}}
\crefrangelabelformat{equation}{(#1) -- (#2)}

\Crefname{equation}{Equation}{Equations}
\Crefname{figure}{Figure}{Figures}
\Crefname{tabular}{Tabular}{Tabulars}

\newcommand{\MethodFull}{Phase-Field/Discontinuity Capturing}
\newcommand{\MethodAbr}{PF/DC}
\newcommand{\Methodfull}{phase-field/discontinuity capturing}

\newcommand{\pfrac}[2]{\frac{\partial #1}{\partial #2}}
\newcolumntype{C}[1]{>{\centering\let\newline\\\arraybackslash\hspace{0pt}}m{#1}}

\newcommand{\keyidea}[1]{%
  \ifshowhighlight
    \textcolor{red}{\textbf{[Key Idea: #1]}}
  \fi
}

\newcommand{\dd}{{\rm d}}

\newtheorem{thm}{Theorem}[section]

\newtheorem{remark}[thm]{Remark}

\begin{document}
\let\printorcid\relax
\def\floatpagepagefraction{1}
\def\textpagefraction{.001}

\shorttitle{PFDC and DVS}    
\shortauthors{Hu, T., Hughes, T.J.R., Scovazzi, G., and Gomez, H.}  
\title[mode = title]{{\MethodFull} operator for direct van der Waals simulation (DVS)}  

\author[Purdue]{Tianyi Hu}
\cormark[1]
\ead{hu450@purdue.edu}

\author[UT]{Thomas J.R. Hughes}

\author[Duke]{Guglielmo Scovazzi}

\author[Purdue]{Hector Gomez}
\ead{hectorgomez@purdue.edu}

\cormark[1]
\cortext[1]{Corresponding author}

\affiliation[Purdue]{
    organization={School of Mechanical Engineering, Purdue University},
    addressline={585 Purdue Mall},
    city={West Lafayette},
    postcode={47906},
    state={Indiana},
    country={USA}
}

\affiliation[UT]{
    organization={Oden Institute for Computational Engineering and Sciences, The University of Texas at Austin},
    addressline={201 East 24th Street},
    city={Austin},
    postcode={78712},
    state={Texas},
    country={USA}
}

\affiliation[Duke]{
    organization={Department of Civil and Environmental Engineering, Duke University},
    addressline={121 Hudson Hall},
    city={Durham},
    postcode={27708},
    state={NC},
    country={USA}
}

\begin{abstract}
    Discontinuity capturing (DC) operators are commonly employed to numerically solve problems involving sharp gradients in the solution. Despite their success, the application of DC operators to the direct van der Waals simulation (DVS) remains challenging. The DVS framework models non-equilibrium phase transitions by admitting interfacial regions in which the derivative of pressure with respect to density is negative. In these regions, we demonstrate that classical DC operators may violate the free energy dissipation law and produce unphysical wave structures. To address this limitation, we propose the {\Methodfull} ({\MethodAbr}) operator. Numerical results show that {\MethodAbr} yields stable and accurate solutions in both bulk fluids and interfacial regions. Finally, we apply the proposed method to simulate cavitating flow over a three-dimensional bluff body, obtaining excellent agreement with experimental data and significant improvements over results produced using classical DC operators.
\end{abstract}



\maketitle


\section{Introduction}\label{sec:intro}

\keyidea{The importance of phase-transforming fluids} Flows with phase transformations are central to a wide range of physical phenomena and engineering applications. These transformations can be triggered by temperature (boiling) or pressure (cavitation) variations \cite{Brennen2013-sy}. The former is widely used in thermal management \cite{Dario2013-zn,Shen2021-ni}, while the latter has broad applications in biotechnologies, such as targeted therapeutic delivery \cite{Husseini2005-ze} and enhanced chemotherapy \cite{Leenhardt2019-zu}. Cavitation also occurs frequently in marine environments, particularly in low-pressure regions induced by rotating propellers. When advected into high-pressure regions, these cavitating bubbles can undergo violent collapse, producing extreme conditions such as temperatures exceeding 5,000 K \cite{Flint1991-vd}, sonoluminescence \cite{Suslick1990-ov}, and strong pressure waves \cite{Johnsen2009-il}. These dynamics can lead to detrimental effects including noise, vibration, and material erosion \cite{Philipp1998-zq}.

\keyidea{Modeling limitations of phase-transforming fluids} Despite their significance, our understanding of phase-transforming flows remains limited, partially due to modeling and computational challenges. Classical models typically describe phase transitions by assuming thermodynamic equilibrium or by introducing empirical mass transfer functions \cite{Koch2016-kb,Osher2001-ow,Gnanaskandan2015-ni}. These mass transfer functions often contain parameters that depend not only on material properties but also on specific flow conditions \cite{Frikha2008-nm}. As a result, they require frequent recalibration, limiting the predictive capability of the models. From a computational perspective, phase-transforming fluids are challenging as they exhibit complex structural features across a wide range of temporal and spatial scales. Moreover, they involve sharp contrasts in density, viscosity, and speed of sound between the liquid and vapor phases, further complicating the development of numerical algorithms.

\keyidea{Development of DVS} A recently proposed computational paradigm, the Direct van der Waals Simulation (DVS) method \cite{Hu2023-zr,Hu2025-by}, enables the simulation of phase-transforming flows with excellent agreement to experimental data of turbulent cavitating flows. Unlike classical methods, DVS does not rely on empirical mass transfer functions to model non-equilibrium thermodynamics at the liquid–vapor interface. Instead, it couples van der Waals’ theory of capillarity \cite{van-der-Waals1979-fk} with the principles  of continuum mechanics \cite{Coleman1963-ob,Gomez2023-xw} to describe the dynamic phase transition process. This coupling also introduces a thermodynamically consistent third-order spatial derivative in the governing equations, enabling DVS to describe surface tension and the propagation of mixed hyperbolic-dispersive wave structures observed in cavitating flows \cite{Pennings2015-va}. However, the mixed-wave structures cannot be decoupled from each other, making the use of classical Godunov-type finite-volume approaches very difficult or even impossible \cite{Dhaouadi2022-os}. To address this challenge, ref. \cite{Hu2025-by} introduced the dispersive Streamline/Upwind Petrov–Galerkin (dispersive-SUPG) method within the finite element framework. Building on the original SUPG method \cite{Brooks1982-og,Shakib1991-ji}, dispersive-SUPG accounts for the characteristics of hyperbolic and dispersive waves, enabling the computation of robust and high-order accurate solutions when either wave type dominates the flow physics.

\keyidea{Need for and Applications of Discontinuity Capturing (DC)} In addition to the mixed hyperbolic–dispersive wave structures, phase-transforming flows often exhibit features with abrupt changes in physical properties, such as shock waves and boundary layers. These regions contain sharp gradients or even discontinuities that cannot be accurately resolved on typical engineering mesh sizes. Such under-resolution leads to spurious oscillations in the numerical solution, which can compromise numerical stability and produce unphysical results. To mitigate this issue, computational methods are often augmented with numerical dissipation to smear out sharp layers and stabilize the solution. In the finite element method (FEM) context, this additional dissipation is formulated as a discontinuity capturing (DC) operator. Originally proposed by \cite{Hughes1986-vj} for the scalar advection equation, the DC operator introduces artificial diffusion aligned with the solution gradient and scaled by the strong-form residual of the governing equations to ensure consistency. This formulation preserves the optimal convergence rate for smooth solutions while providing stability in the presence of sharp gradient. The DC operator was later generalized to advective–diffusive systems in \cite{Hughes1986-ps} and extended to the compressible Euler and Navier–Stokes equations in \cite{Shakib1991-ji, Le-Beau1993-ip}. Since then, numerous variants have been developed and successfully applied to problems involving supersonic flows \cite{Tezduyar2006-vo, Codoni2021-fi, Rajanna2022-so}, shock hydrodynamics \cite{Zeng2016-zy, Scovazzi2007-ts}, and biomedical applications \cite{Bazilevs2007-cl, Wang2023-wl}.

\keyidea{Motivation for a New Discontinuity Capturing (DC) Operator} Although DC operators have demonstrated success in gas dynamics, they cannot be directly applied to the DVS framework for two key reasons. First, DVS is capable of describing coexisting vapor and liquid phases with markedly different characteristics: while the vapor phase is highly compressible and often features shock waves, the liquid phase is nearly incompressible and typically does not need any discontinuity capturing. Although classical DC operators perform well in the vapor phase, they do not naturally vanish in the liquid phase, leading to over-dissipation and degradation of solution accuracy. Second, and more critically, the liquid-vapor interfacial region may exhibit sharp density gradients with wave structures that are generally inadmissible in conventional compressible flow, and thus fall outside the design considerations of classical DC operators. In fact, we demonstrate that in these interfacial regions, classical DC operators can violate the free energy dissipation law, generate artificial wave structures, and lead to over-dissipation of interfacial dynamics.

\keyidea{Proposed PFDC Method and Paper Structure}
In this paper, we propose a new stabilization technique, {\Methodfull} ({\MethodAbr}), to fundamentally address the two key challenges posed by the application of classical DC operators in the DVS framework. We demonstrate that the {\MethodAbr} operator (1) achieves optimal rate of convergence in both interfacial and bulk regions and (2) satisfies the free energy dissipation law at liquid-vapor interfaces. The paper is organized as follows. In \cref{sec:NSK}, we derive the governing equations of the DVS framework. In \cref{sec:DC_PFDC}, we review the design principles of classical DC operators and highlight their limitations in the context of DVS. We then introduce the formulation of the {\MethodAbr} operator and explain the design philosophy. In \cref{sec:PFDC_DVS}, we describe the complete numerical method and solution procedure. Finally, we conduct a series of numerical experiments to compare the performance of classical DC and {\MethodAbr} operators. Our results show that {\MethodAbr} significantly improves both accuracy and stability, yielding much better agreement with experimental observations.

\section{The isothermal Navier-Stokes-Korteweg equations} 
\label{sec:NSK}

The DVS framework relies on the Navier–Stokes–Korteweg (NSK) equations. We begin by presenting a brief derivation of the isothermal form of the model. Next, we discuss the choice of equation of state (EoS) and the associated thermodynamic properties. We then describe the interfacial dynamics and the corresponding equilibrium conditions. Finally, we derive the free energy dissipation law that governs the system's evolution toward equilibrium.

\subsection{Model derivation}

\keyidea{Introduce the conservation law (most general)} The derivation of isothermal NSK equations starts from conservation of mass,
\begin{equation}
    \pfrac{\rho}{t} + \nabla \cdot \left(\rho \bm{u}\right) = 0,
    \label{eqn:CoM}
\end{equation}
conservation of linear momentum,
\begin{equation}
    \pfrac{\left(\rho \bm{u}\right)}{t} + \nabla \cdot \left(\rho \bm{u}\otimes \bm{u}\right) = \nabla \cdot \bm{\sigma},
    \label{eqn:CoLM}
\end{equation}
and conservation of angular momentum,
\begin{equation}
    \bm{\sigma} = \bm{\sigma}^\intercal.
    \label{eqn:CoAM}
\end{equation}
Here, $t$ represents time, $\rho$ is the fluid density, $\bm{u}$ is the velocity, and $\bm{\sigma}$ is the Cauchy stress tensor. 
\keyidea{Introduce constitution relationship for NSK and Coleman-Noll (less general)} The isothermal NSK equations are developed by postulating a non-local free energy per unit volume \cite{van-der-Waals1979-fk}
\begin{equation}
    \Psi(\rho,\rho \bm{u}) = \rho f(\rho) + \frac{1}{2}\lambda F |\nabla \rho|^2 + \frac{1}{2} \frac{|\rho\bm{u}|^2}{\rho},
    \label{eqn:free_energy}
\end{equation}
where $f(\rho)$ is the isothermal Helmholtz free energy per unit mass, while $\lambda$ and $F$ are constants that control interfacial energy and interface thickness, respectively. Using the Coleman-Noll approach \cite{Coleman1963-ob,Dunn1985-xc,Gomez2023-xw}, we can find the constitutive relations for $\bm{\sigma}$ that ensure the total free energy within the domain $\Omega$,
\begin{equation}
    \mathcal{E}(\rho,\rho \bm{u}) = \int_{\Omega} \Psi(\rho,\rho \bm{u}) \dd \Omega,
    \label{eqn:total_energy}
\end{equation}
decreases with time for any arbitrary process that satisfies \cref{eqn:CoM,eqn:CoLM,eqn:CoAM}. A choice of $\bm{\sigma}$ compatible with free energy dissipation is: 
\begin{equation}
    \bm{\sigma} = -p\bm{I} + \overline{\mu}\left(\rho\right) \bm{D} + \bm{\zeta},
    \label{eqn:cauchy_stress_tensor}
\end{equation}
where $p = \rho^2 \frac{ d f(\rho)}{d \rho}$ is the thermodynamic pressure, $\bm{I}$ is the identity tensor and viscosity coefficient $\overline{\mu}\left(\rho\right)$ is a positive function of fluid density; see \cite{Kestin1984-en,Hu2025-by}. The deviatoric strain rate tensor is
\begin{equation}
     \bm{D} =  \nabla \bm{u} + \nabla \bm{u}^\intercal - \frac{2}{3}\nabla \cdot \bm{u} \bm{I}.
    \label{eqn:strain_rate_tensor}
\end{equation}
The Korteweg stress tensor $\bm{\zeta}$ accounts for the interfacial stresses and can be written as  
\begin{equation}
    \bm{\zeta} = \lambda F \left[\left(\rho \Delta \rho + \frac{1}{2}\vert\nabla \rho\vert^2\right)\bm{I} - \nabla \rho \otimes \nabla \rho \right].
    \label{eqn:korteweg_stress}
\end{equation}
\keyidea{Introduce equation in compact form (less general)} The isothermal NSK equations can also be compactly written as 
\begin{equation}
    \bm{{U}}_{,t} + \bm{{F}}^{\rm hype}_{i,i} + \bm{{F}}^{\rm disp} = \bm{{F}}^{\rm diff}_{i,i},
    \label{eqn:DVS_CompactForm}
\end{equation}
where an inferior comma denotes partial differentiation (e.g., $\bm{{U}}_{,t}=\partial\bm{U}/\partial t$), and repeated indices indicate summation over the spatial dimensions (e.g., $\bm{{F}}^{\rm hype}_{i,i}=\sum_{i=1}^{d}\partial\bm{{F}}^{\rm hype}_{i}/\partial x_i$, where the subscript $i$ denotes the $i$th Cartesian coordinate component and $d$ is the number of spatial dimensions). The vector 
\begin{equation}
    \bm{{U}} = \left[\rho,\rho u_1,\rho u_2,\rho u_3\right]^\intercal
    \label{eqn:ConservedVariable}
\end{equation}
contains the conserved variables. The vectors $\bm{{F}}^{\rm hype}_i$, $\bm{{F}}^{\rm diff}_i$, and $\bm{{F}}^{\rm disp}$ represent the hyperbolic fluxes \footnote{For the NSK equations, the term ``hyperbolic flux'' is only accurate when $d p/ d \rho > 0$. However, we retain the use of ``hyperbolic'' to maintain consistency with the literature.}, the diffusive fluxes, and the dispersive stress contribution, respectively, and they are defined as follows:
\begin{equation}
    \bm{{F}}^{\rm hype}_i = \bm{{F}}^{\rm {hype \textbackslash p}}_i + \bm{{F}}^{p}_i = 
    \begin{bmatrix} 
      \rho u_i \\ 
      \rho u_1 u_i\\ 
      \rho u_2 u_i\\
      \rho u_3 u_i
    \end{bmatrix}
    + 
    \begin{bmatrix} 
      0 \\ 
      p\delta_{1i}\\ 
      p\delta_{2i}\\
      p\delta_{3i}
    \end{bmatrix}
    \mbox{, }
    \bm{{F}}^{\rm diff}_i = \overline{\mu}(\rho)
    \begin{bmatrix} 
      0 \\ 
      \bm{D}_{1i} \\ 
      \bm{D}_{2i} \\ 
      \bm{D}_{3i} 
    \end{bmatrix}
    \mbox{, }
    \bm{{F}}^{\rm disp} = 
    \begin{bmatrix} 
      0 \\ 
      -\lambda F \rho \Delta \rho_{,1}\\ 
      -\lambda F \rho \Delta \rho_{,2}\\
      -\lambda F \rho \Delta \rho_{,3}
    \end{bmatrix}.
    \label{eqn:DVS_CompactForm_Fluxes}
  \end{equation}
In the equation above, $\delta_{ij}$ is the Kronecker delta and we have exploited the identity $\nabla \cdot \bm{\zeta} = \lambda F\rho \nabla \left(\Delta \rho\right)$ \cite{Gomez2010-kt}.

\subsection{Equation of state and thermodynamics}
\keyidea{Brief discussion about EoS} An equation of state (EoS) is a thermodynamic equation that relates the macroscopic state variables of a fluid. A well-known example is the ideal gas EoS, which assumes that particles have no volume and there are no intermolecular forces between them, leading to a monotonic pressure-density relationship. Although widely used in compressible flow, the ideal gas law fails to capture the behavior of phase-transforming fluids. In particular, it cannot predict the coexistence of liquid-vapor phases or the phase transition process. 

\begin{figure}[htbp]
    \centering
    \includegraphics[keepaspectratio=true,width=\linewidth]{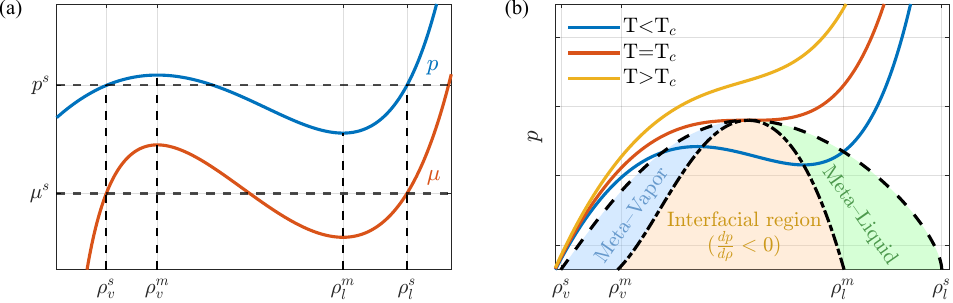}
    \caption{(a) Pressure and chemical potential as functions of density for a cubic equation of state (EoS) below the critical condition. (b) Pressure  as a function of density at temperatures below, at, and above the critical temperature. The dashed and dash-dot lines in (b) mark the predicted phase envelope and the interfacial region, respectively.}
    \label{fig:EoS}
\end{figure}

\keyidea{Discussion about cubic EoS and it's characteristic for multiphase flow} To address this limitation, van der Waals (vdW) proposed a modified EoS that accounts for the effects of finite molecular size and intermolecular forces~\cite{van-der-Waals1873-cc}. This framework leads to a class of cubic EoS that may admit multiple densities for a given thermodynamic state; see \cref{fig:EoS}(a). This feature permits the coexistence of stable liquid and vapor phases. The corresponding equilibrium conditions are obtained by solving the following problem~\cite{Kunz2012-qe,Bell2018-sd}: For a given temperature $T$ below the critical point, find a pair of liquid and vapor densities, denoted by $\rho^s_l$ and $\rho^s_v$, that satisfy pressure equilibrium
\begin{equation}
   p^s \left(T\right) = p\left(\rho_l^s(T)\right) = p\left(\rho_v^s(T)\right), 
   \label{eqn:PressureEqui}
\end{equation}
and chemical potential equilibrium
\begin{equation}
   \mu^s\left(T\right) = \mu\left(\rho_l^s(T)\right) = \mu\left(\rho_v^s(T)\right), 
   \label{eqn:ChemEqui}
\end{equation}
where chemical potential is given by $\mu = f(\rho) + \rho \frac{d f(\rho)}{d \rho}$, and $p^s$ and $\mu^s$ denote the saturation pressure and chemical potential, respectively. When the saturation conditions exist, a cubic EoS also identifies an interfacial region within the density range $\rho\in(\rho_{v}^m,\rho_{l}^m)$, where the spinodal densities $\rho_{v}^m$ and $\rho_{l}^m$ are the roots of $\frac{d p}{d \rho} = \frac{d \mu}{d \rho} = 0$ that are closest to $\rho_v^s$ and $\rho_l^s$. Within this region, the fluid is thermodynamically unstable; any perturbation can lead to spontaneous phase separation~\cite{Cahn1965-vd}. The metastable vapor and liquid lie between the spinodal and saturation densities, i.e., in the ranges $\rho \in (\rho_v^s, \rho_v^m)$ and $\rho \in (\rho_l^m, \rho_l^s)$, respectively. In the metastable region, the fluid is mechanically stable, but phase separation can occur in the presence of nucleation sites or sufficiently large perturbations. As temperature increases toward the critical temperature $T_c$, the saturation densities $\rho_l^s$ and $\rho_v^s$ approach each other, and both the metastable and interfacial regions shrink. At the critical point ($T = T_c$), the two saturation densities coincide and there no longer exist distinct vapor and liquid phases. For $T > T_c$, the fluid enters the supercritical regime, and we have a monotonic pressure-density relationship similar to the ideal gas law; see \cref{fig:EoS}(b). Thus, the class of cubic EoS provides a unified framework for capturing phase separation, critical phenomena, and the subcritical-supercritical transition, making them well-suited for modeling multiphase flows.

\keyidea{Discussion limitation about cubic EoS} Since the original vdW EoS was introduced, numerous modifications have been developed to enhance its predictive capabilities. Among these, the Soave–Redlich–Kwong (SRK) \cite{Soave1972-zh} and Peng–Robinson (PR) \cite{Peng1976-lw} EoS are the most widely used for practical applications. While these models are capable of predicting the phase behavior of many nonpolar and mildly polar fluids, the underlying structure of cubic EoS does not account for association effects or specific bonding mechanisms, which are critical in strongly polar substances such as water~\cite{Kontogeorgis2010-ta}. These effects become more pronounced as fluids move away from the critical point, leading to significant deviations between predicted and actual phase envelopes.

\begin{figure}[htbp]
    \centering
    \includegraphics[keepaspectratio=true,width=\linewidth]{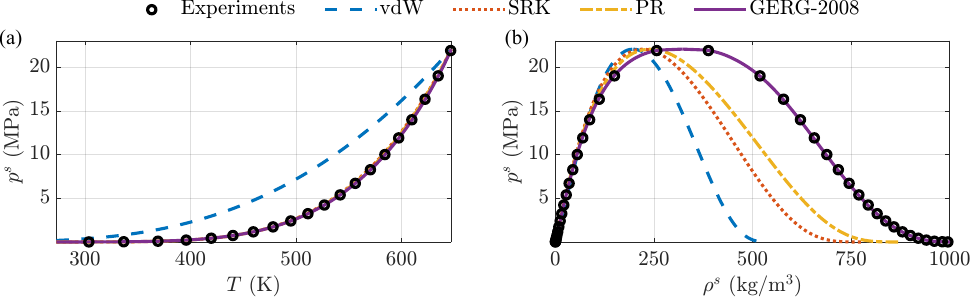}
    \caption{(a) Saturation pressure ($p^s$) as a function of temperature ($T$), and (b) corresponding phase envelopes for water predicted by various equations of state (EoS). Markers represent experimental data from \cite{Linstrom1997-pt}. The results show that the GERG-2008 EoS provides the best agreement with experimental measurements.}
    \label{fig:Saturation}
\end{figure}

\keyidea{Discussion multi parameter EoS and it's advantages} In recent decades, a new class of multi-parameter equations of state has emerged. These models are formulated explicitly in terms of the Helmholtz free energy, with parameters obtained through empirical fitting to high-accuracy experimental data. As a result, they provide excellent agreement with measured phase behavior across a broad range of conditions. Among them, the GERG–2008 EoS \cite{Kunz2012-qe} covers 21 natural gas components and remains valid over a wide range of temperatures and pressures, making it well-suited for accurate modeling of cavitation and boiling problems, where extreme fluid conditions are often encountered \cite{Magaletti2015-gi}. In \cref{fig:Saturation}, we present the predicted (a) water saturation pressure as a function of temperature, and (b) the water phase envelopes using various EoS. The results are obtained by iteratively solving \cref{eqn:PressureEqui,eqn:ChemEqui} for each EoS~\cite{Sandler2017-kn}. In the same figure, we also plot experimental water data from NIST~\cite{Linstrom1997-pt}. We observe that the vdW EoS yields inaccurate saturation pressures and phase envelopes. While the SRK and PR EoS significantly improve the saturation pressure prediction, the resulting liquid density remains much lower than the experimental value. In contrast, the GERG–2008 EoS obtains excellent agreement with the experimental data.

\keyidea{Discussion challenges of multi parameter EoS and solution} Although the GERG-2008 EoS obtains superior accuracy compared to the class of cubic EoS, it is not admissible for models based on density gradient theory \cite{Wilhelmsen2017-cq} such as \cref{eqn:DVS_CompactForm}. Under this theory, $\frac{d p}{d \rho}$ must remain strictly negative within the interfacial region, a condition naturally satisfied by cubic EoS. However, the GERG-2008 EoS originates from equilibrium thermodynamics by fitting data of bulk and metastable fluid. This leads to the potential unphysical oscillations in the interfacial region. To address this issue, a piecewise reconstruction of GERG-2008 has been proposed \cite{Magaletti2021-qv,Hu2025-by}. Such modification preserves the accurate prediction of bulk and metastable properties of the original GERG-2008 EoS, while ensuring thermodynamic admissibility by mimicking the shape of cubic EoS within the interfacial region. In this paper, we adopt the modification in \cite{Hu2025-by}.

\subsection{Interfacial dynamics and equilibrium condition}

\keyidea{Issue of NS at interfacial region} Conventional compressible flow models do not admit the existence of an interfacial region where $\frac{d p}{d \rho} < 0$. In these regions, the isentropic form of the governing equations becomes a mixed hyperbolic-elliptic type, resulting in an ill-posed problem \cite{Menikoff1989-dj}. 

\keyidea{How NSK can admit interfacial region} In contrast, the NSK system incorporates the Korteweg stress, \cref{eqn:korteweg_stress}, ensuring well-posedness even when the EoS is nonconvex \cite{Danchin2001-yw}. At the liquid-vapor interface, the negative $\frac{d p}{d \rho}$ is associated with anti-diffusive behavior, which can induce exponential growth of perturbations. As the gradient of the solution increases, the higher-order Korteweg stress becomes dominant and introduces dispersion, providing an additional transport mechanism for perturbations. Because the group velocity scales quadratically with the mode frequency, the dispersive effect confines low-frequency modes locally, allowing them to continuously grow in a controlled manner. Meanwhile, high-frequency modes are transported away from the interfacial region, preventing the unbounded growth of perturbations. Eventually, the low-frequency modes develop into a stable liquid-vapor interface, while the high-frequency modes propagate into the bulk region and form an oscillatory wave train. In the presence of viscosity, these wave trains are progressively damped, ultimately yielding a non-oscillatory equilibrium interface between the liquid and vapor phases. 

\begin{figure}[htbp]
    \centering
    \includegraphics[keepaspectratio=true,width=\linewidth]{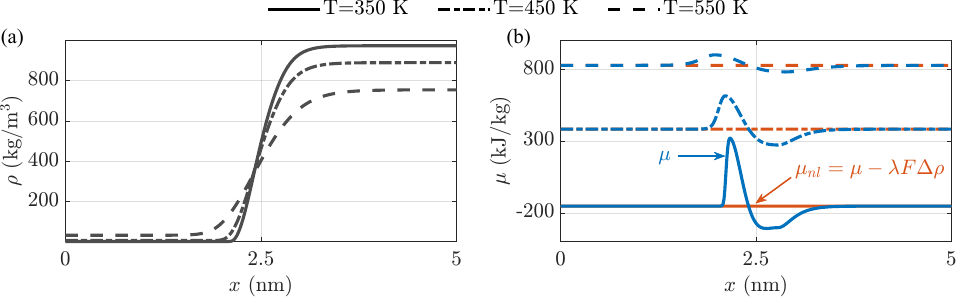}
    \caption{(a) Equilibrium liquid–vapor interface profiles along the spatial coordinate $x$ at temperatures $T = 350$, $450$, and $550$ K. (b) Corresponding profiles of the chemical potential $\mu$ and the non-local chemical potential $\mu_{\rm nl}$.}
    \label{fig:EquilibriumProfile}
\end{figure}

\keyidea{Euler-Lagrange Condition} The equilibrium condition can also be directly obtained by minimizing the total free energy \cref{eqn:total_energy} under the constraints of mass conservation and  $\bm{u} = \bm{0}$. This yields the Euler-Lagrange condition \cite{Magaletti2015-gi}
\begin{equation}
\mu_{\rm nl} = \mu - \lambda F  \Delta \rho = \mathcal{L},
\label{eqn:EulerLagrange}
\end{equation}
where $\mu_{\rm nl}$ denotes the non-local chemical potential, and the constant $\mathcal{L}$ is the Lagrange multiplier associated with mass conservation. At a given temperature, we can obtain the equilibrium interface profile by solving \cref{eqn:EulerLagrange} using stabilized density gradient theory \cite{Mu2017-wn}. In \cref{fig:EquilibriumProfile}, we plot (a) the liquid–vapor interface profile and (b) the corresponding $\mu$ and $\mu_{\rm nl}$ along the spatial coordinate $x$ at temperatures $T = 350$, $450$, and $550$ K. We observe that $\mu$ varies with $x$ in equilibrium conditions, while $\mu_{\rm nl}$ remains spatially uniform and equal to the Lagrange multiplier $\mathcal{L}$. This property will be exploited when developing numerical algorithms; see \cref{sec:PFDC}. Furthermore, we notice that as the temperature decreases, the density {change} across the interface increases, while the interface thickness becomes progressively thinner. This behavior presents significant challenges for the numerical simulation of strongly undercritical fluids and highlights the need for robust numerical methods.

\subsection{Free energy dissipation law}
\label{sec:EnergyDissipation}
\keyidea{Introduce functional entropy variable} It is also of interest to derive the exact free energy dissipation relation that governs the system’s evolution toward equilibrium. The derivation begins with the definition of the functional entropy variable $\bm{V}$ \cite{Liu2013-kl, Liu2015-mu}. For the isothermal NSK equations, $\bm{V}$ is obtained by taking the functional derivative of the total free energy $\mathcal{E}$ with respect to the conserved variables $\bm{U}$:
\begin{equation}
    \bm{V} = \frac{\delta \mathcal{E}}{\delta \bm{U}} = \left[\mu_{\rm nl} - \frac{|\bm{u}|^2}{2},u_1,u_2,u_3\right]^\intercal.
    \label{eqn:EntropyVariable}
\end{equation}
\keyidea{Derivation of energy dissipation law} Assuming the solution $\left(\rho, \rho\bm{u}\right)$ is sufficiently smooth, we contract the NSK equations \cref{eqn:DVS_CompactForm} with the entropy variable $\bm{V}$ and integrate over the fixed domain $\Omega$:
\begin{equation}
    \int_\Omega {\bm{V} \cdot \left[\bm{{U}}_{,t} + \bm{{F}}^{\rm hype}_{i,i} + \bm{{F}}^{\rm disp}\right]} {\rm d}\Omega = \int_\Omega \bm{V} \cdot \bm{{F}}^{\rm diff}_{i,i} {\rm d}\Omega.
    \label{eqn:EnergyDissipation_Inter1}
\end{equation}
Substituting \cref{eqn:EntropyVariable,eqn:ConservedVariable,eqn:DVS_CompactForm_Fluxes} into \cref{eqn:EnergyDissipation_Inter1} and expanding the terms, we obtain:
\begin{align}
    \int_\Omega \left(\mu_{\rm nl} - \frac{|\bm{u}|^2}{2}\right) \left[\pfrac{\rho}{t} + \nabla \cdot \left(\rho \bm{u}\right)\right] {\rm d}\Omega &+ \int_\Omega \bm{u} \cdot \left[\pfrac{\left(\rho \bm{u}\right)}{t} + \nabla \cdot \left(\rho \bm{u}\otimes \bm{u}\right)  + \nabla p -  \lambda F \rho \nabla  \Delta \rho \right] {\rm d}\Omega \nonumber \\
    &= \int_\Omega \bm{u} \cdot \left[\nabla \cdot \left(\overline{\mu}\left(\rho\right) \bm{D}\right)\right] {\rm d}\Omega.
    \label{eqn:EnergyDissipation_Inter2}
\end{align}
Exploiting the identities
\begin{equation}
\frac{d \mathcal{E}}{dt} = \int_\Omega  \left[\frac{\delta \mathcal{E}}{\delta \rho} \pfrac{\rho}{t} + \int_\Omega \frac{\delta \mathcal{E}}{\delta \left(\rho \bm{u}\right)} \cdot \pfrac{\left(\rho \bm{u}\right)}{t}\right]{\rm d}\Omega  = \int_\Omega  \left[\left(\mu_{\rm nl} - \frac{|\bm{u}|^2}{2}\right)\pfrac{\rho}{t} + \bm{u} \cdot \pfrac{\left(\rho \bm{u}\right)}{t}\right] {\rm d}\Omega,
\label{eqn:EnergyDissipation_Property1}
\end{equation}
and 
\begin{equation}
\nabla p - \lambda F \rho \nabla \Delta \rho = \rho \nabla\left(\mu - \lambda F \Delta\rho\right) = \rho \nabla \mu_{\rm nl} ,
\label{eqn:EnergyDissipation_Property2}
\end{equation}
we can rewrite \cref{eqn:EnergyDissipation_Inter2} as:
\begin{equation}
\frac{d \mathcal{E}}{dt} + \int_\Omega \left[\nabla \cdot \left(\rho \bm{u}\right) \left(\mu_{\rm nl} - \frac{|\bm{u}|^2}{2}\right) + \bm{u} \cdot \nabla \cdot \left(\rho \bm{u} \otimes \bm{u}\right) + \bm{u} \cdot \rho \nabla \mu_{\rm nl} \right] {\rm d}\Omega = \int_\Omega \bm{u} \cdot \nabla \cdot \left(\overline{\mu}(\rho) \bm{D}\right) {\rm d}\Omega.
\label{eqn:EnergyDissipation_Inter3}
\end{equation}
We perform integration by parts on \cref{eqn:EnergyDissipation_Inter3} and apply periodic boundary conditions on all boundaries, yielding:
\begin{equation}
\frac{d \mathcal{E}}{dt} + \int_\Omega \left[ -\rho \bm{u} \cdot \nabla \mu_{\rm nl} + \frac{1}{2} \rho \bm{u} \cdot \nabla \left(|\bm{u}|^2\right) - \nabla \bm{u} : \rho \bm{u} \otimes \bm{u} + \bm{u} \cdot \rho \nabla \mu_{\rm nl} \right] {\rm d}\Omega = - \int_\Omega \nabla \bm{u} : \overline{\mu}(\rho) \bm{D} {\rm d}\Omega.
\label{eqn:EnergyDissipation_Inter4}
\end{equation}
Note that the identity $\bm{u} \cdot \nabla \left(|\bm{u}|^2\right) = 2 \nabla \bm{u} : \bm{u} \otimes \bm{u}$ implies that all terms within the integrand on the left-hand side of \cref{eqn:EnergyDissipation_Inter4} cancel each other out. Finally, we obtain the following free energy dissipation law:
\begin{equation}
\frac{d}{dt} \mathcal{E}(\rho, \rho \bm{u}) = - \int_\Omega \nabla \bm{u} : \overline{\mu}(\rho) \bm{D} {\rm d}\Omega \leq 0.
\label{eqn:EnergyDissipation}
\end{equation}
This result shows that for fluids with positive viscosity, the isothermal NSK equations dissipate total free energy over time for any arbitrary process, thereby ensuring thermodynamic stability.

\section{Design of the discontinuity capturing operator}
\label{sec:DC_PFDC}

The solution of the compressible flow equations may contain sharp gradients or discontinuities. In such cases, direct numerical discretization of \cref{eqn:DVS_CompactForm} can result in unphysical oscillations in the numerical solution or lead to divergence of the algorithm. Within the context of the finite element method (FEM), it is common to augment \cref{eqn:DVS_CompactForm} with a discontinuity-capturing (DC) operator:
\begin{equation}
    \bm{{U}}_{,t} + \bm{{F}}^{\rm hype}_{i,i} + \bm{{F}}^{\rm disp} = \bm{{F}}^{\rm diff}_{i,i} + \bm{{F}}^{\rm DC}_{i,i}.
    \label{eqn:DVS_CompactForm_Mod}
\end{equation}
To preserve consistency, the added flux $\bm{{F}}^{\rm DC}_{i}$ must be designed such that any solution of \cref{eqn:DVS_CompactForm} is also a solution of \cref{eqn:DVS_CompactForm_Mod}. This requirement is often satisfied by linking the magnitude of $\bm{{F}}^{\rm DC}_i$ to the strong-form residual of \cref{eqn:DVS_CompactForm},
\begin{equation}
    {\bf Res}\left(\bm{U}\right) = \bm{{U}}_{,t} + \bm{{F}}^{\rm hype}_{i,i} + \bm{{F}}^{\rm disp} - \bm{{F}}^{\rm diff}_{i,i},
    \label{eqn:StrongFormResidual}
\end{equation}
such that $\bm{{F}}^{\rm DC}_i$ vanishes as ${\bf Res}\left(\bm{U}\right)$ tends to zero. 

In this section, we first review classical DC operator designs and their recent modifications for phase-transforming fluids. We then examine the limitations of these methods when applied to the NSK equations. Finally, we present the formulation and design philosophy of the {\Methodfull} operator, demonstrating that it fundamentally resolves the aforementioned challenges.

\subsection{The classical DC operator}

\keyidea{Present the formulation of classical DC operator.} Following \cite{Shakib1991-ji,Codoni2021-fi}, we define the classical DC operator as
\begin{equation}
    \bm{{F}}^{\rm DC}_i = \kappa_{\rm DC} \bm{U}_{,i},
    \label{eqn:DC_classical}
\end{equation}
where the non-negative scalar $\kappa_{\rm DC}$ determines the intensity of the DC operator, and $\bm{U}_{,i}$ determines the direction in which the DC operator is applied. The value of $\kappa_{\rm DC}$ is defined as the minimum of two contributions:
\begin{equation}
    \kappa_{\rm DC} = \min \left(\tilde{\kappa}_{\rm DC}, \hat{\kappa}_{\rm DC}\right),
    \label{eqn:kappa_DC}
\end{equation} 
where 
\begin{equation}
    \tilde{\kappa}_{\rm DC} = C_{\rm DC}\left(\frac{{\bf Res}^{\intercal} \bm{\mathcal{N}} {\bf Res}}{G_{ij} \bm{U}_{,i}^{\intercal} \bm{\mathcal{N}}\bm{U}_{,j}}\right)^{\frac{1}{2}},
  \label{eqn:kappa_DC_tilde}
\end{equation}
and
\begin{equation}
  \hat{\kappa}_{\rm DC} = \left[\bm{u}\cdot \bm{G}^{-1}\bm{u} + p_{,\rho}^{+}\mbox{tr}\left(\bm{G}^{-1}\right)\right]^{1/2}.
  \label{eqn:kappa_DC_hat}
\end{equation}
Here, $C_{\rm DC}$ is an $\mathcal{O}(1)$ constant\footnote{Unless otherwise stated, we set $C_{\rm DC} = 1$ in the numerical simulations.}, and the quantity $p_{,\rho}^+ = \max\left(0, \frac{dp}{d\rho}\right)$ represents the positive part of the pressure derivative with respect to density. The matrix $G_{ij}$ denotes the components of the element metric tensor $\bm{G}$, defined as
\begin{equation}
  G_{ij} = \sum _{k=1}^d \pfrac{\xi_k}{x_i}\pfrac{\xi_k}{x_j},
  \label{eqn:GeometricTensor}
\end{equation}
where $\bm{x}\left(\bm{\xi}\right)$ is the element isoparametric mapping. The  tensor $\bm{G}$ describes the local element deformation and is related to the element mesh size and anisotropy. The semi-positive definite matrix $\bm{\mathcal{N}}$ ensures the dimensional consistency when contracting strong-form residual vector. We define it as
\begin{equation}
    \bm{\mathcal{N}} = 
    \begin{bmatrix}
        \left(p_{,\rho}^{+}\right)^2 & 0 & 0 & 0 \\
        0 & |\bm{u}|^2 & 0 & 0\\
        0 & 0 & |\bm{u}|^2 & 0\\
        0 & 0 & 0 & |\bm{u}|^2\\
    \end{bmatrix}.
\end{equation}
When the solution is well-resolved by the discretization, the strong-form residual remains small, causing $\tilde{\kappa}_{\rm DC}$ to vanish and preserving the accuracy of the numerical solution. In under-resolved regions, $\tilde{\kappa}_{\rm DC}$ scales linearly with the mesh size, thereby enhancing the robustness of the scheme. While generally well-behaved, the formulation in \cref{eqn:kappa_DC_tilde} involves division by the solution gradient, which can cause $\tilde{\kappa}_{\rm DC}$ to become locally large in regions with very small gradients. To mitigate this issue, we use multi-dimensional generalization of the upwind viscosity, $\hat{\kappa}_{\rm DC}$, as an upper bound of numerical dissipation. This upper limit suppresses local spikes in $\tilde{\kappa}_{\rm DC}$ and reduces nonlinearity in the evaluation of the DC operator, thereby improving the convergence of the overall numerical scheme.

\keyidea{Discuss the issue of classical DC} However, when applied to the NSK equations, numerical results suggest that the classical DC operator in \cref{eqn:DC_classical} can (1) overdissipate flow features in the liquid phase and (2) induce spurious oscillations near the interfacial region. Both issues can be understood from a free energy dissipation perspective. Following the framework in \cref{sec:EnergyDissipation}, we contract the functional entropy variable $\bm{V}$ with the modified system \cref{eqn:DVS_CompactForm_Mod} and integrate over the domain, assuming periodic boundary conditions\footnote{When deriving the modified free energy dissipation relation, we assume $\kappa_{\rm DC}$ is sufficiently smooth for analytical convenience.}. This yields the modified free energy dissipation law:
\begin{equation}
\frac{d}{dt} \mathcal{E}(\rho, \rho \bm{u}) = - \int_\Omega \nabla \bm{u} : \overline{\mu}(\rho) \bm{D} {\rm d}\Omega - \int_\Omega \kappa_{\rm DC} \left[ \frac{1}{\rho}\frac{dp}{d\rho}|\nabla \rho|^2 - \lambda F \nabla\Delta \rho \cdot \nabla \rho + \rho \nabla\bm{u}:\nabla\bm{u} \right]{\rm d}\Omega.
\label{eqn:EnergyDissipation_ClassicalDC}
\end{equation}
The second term on the right-hand side of \cref{eqn:EnergyDissipation_ClassicalDC} represents the contribution of the DC operator. We first consider the case where the solution consists purely of bulk liquid. In the incompressible limit, a DC operator is not required. However, since the strong-form residual does not analytically vanish in the liquid phase, $\kappa_{\rm DC}$ remains finite. As a result, even small gradients in the density or velocity fields can cause unnecessary artificial dissipation, reducing solution accuracy. We then consider the scenario where the solution contains both liquid and vapor phases. From \cref{eqn:EnergyDissipation_ClassicalDC}, we observe that the last term $\rho \nabla \bm{u}:\nabla \bm{u}$ is non-negative and is thus compatible with the free energy dissipation law. In contrast, the combination of $\frac{1}{\rho} \frac{d p}{d\rho} |\nabla \rho|^2 - \lambda F \nabla \Delta \rho \cdot \nabla \rho$ can become negative in the interfacial region, potentially leading to a net increase in the total free energy. This violation of the free energy dissipation law may result in unphysical oscillations and artificial solution features not present in the original NSK equations.

\subsection{The DC operator with compressibility scaling}
\keyidea{Discuss DC with scaling and it's issue} Recent studies~\cite{Hu2023-zr, Hu2025-by} have attempted to address the aforementioned issues associated with applying the classical DC operator to the numerical solution of the NSK equations. In these works, the authors proposed a modified operator:
\begin{equation}
    \bm{{F}}^{\rm DC}_i = \eta \kappa_{\rm DC} \bm{U}_{,i},
    \label{eqn:DC_classical_wScale}
\end{equation}
where $\eta \in \left[0,1\right]$ is a scaling factor that adjusts the intensity of the DC operator based on the local compressibility of the fluid. Following \cite{Hu2025-by}, $\eta$ is defined as 
\begin{equation}
    \eta = \frac{\left(M_{\rm iso} / M_{\rm iso,r}\right)^2}{\left(M_{\rm iso} / M_{\rm iso,r}\right)^2 + 1},
    \label{eqn:DC_Scale}
\end{equation}
where $M_{\rm iso} = |\bm{u}| / \sqrt{p_{,\rho}^{+}}$ and $M_{\rm iso,r} = 1/3$. When $M_{\rm iso} \ll M_{\rm iso,r}$, the flow is nearly incompressible and $\eta$ approaches zero, effectively suppressing the DC operator to minimize numerical dissipation. As $M_{\rm iso}$ increases beyond $M_{\rm iso,r}$, the flow becomes increasingly compressible and $\eta$ approaches one quadratically, activating the DC operator to enhance algorithm robustness. Numerical results suggest that this simple modification improves the solution accuracy in the bulk liquid. However, it does not resolve the violation of the free energy dissipation law in the interfacial region. In particular, when $\frac{dp}{d\rho} < 0$, any nonzero velocity can cause $\eta$ to approach one. Thus, the scaled operator in \cref{eqn:DC_classical_wScale} often behaves identically to the original operator in \cref{eqn:DC_classical} when $\frac{dp}{d\rho}< 0$, continuing to generate unphysical oscillations.

\subsection{The {\Methodfull} operator}\label{sec:PFDC}
\keyidea{Present the idea and design for PFDC} It is evident that a new DC operator with specialized treatment of the interfacial region is necessary. To resolve the aforementioned challenges, {we further exploit the free-energy dissipation law and} propose the following design
\begin{equation}
    \bm{{F}}^{\rm DC}_i = \eta \kappa_{\rm DC}^{u} \bm{F}_i^{{\rm DC},u} + 
    \begin{cases}
        \eta \kappa_{\rm DC}^{\rho} \bm{F}_i^{{\rm DC},\rho}, & \text{if } \ \frac{dp}{d\rho} > 0 \\[1.5ex]
        \eta \kappa_{\rm DC}^{\mu} \bm{F}_i^{{\rm DC},\mu}, & \text{otherwise} 
    \end{cases}
    \label{eqn:PFDC}
\end{equation}
where $\kappa_{\rm DC}^{u}$, $\kappa_{\rm DC}^{\rho}$, and $\kappa_{\rm DC}^{\mu}$ are non-negative coefficients that control the strength of the corresponding DC fluxes. This new formulation introduces three distinct flux components:
\begin{equation}
    \bm{F}_i^{{\rm DC},u} =  \begin{bmatrix} 
        0 \\ 
        \rho \bm{D}_{1i} \\ 
        \rho \bm{D}_{2i} \\ 
        \rho \bm{D}_{3i} 
    \end{bmatrix}
        , \; 
    \bm{F}_i^{{\rm DC},\rho} =  \begin{bmatrix} 
        \rho_{,i}\\ 
        u_1 \rho_{,i} \\ 
        u_2 \rho_{,i} \\ 
        u_3 \rho_{,i}
    \end{bmatrix}
        , \; 
    \bm{F}_i^{{\rm DC},\mu} =  \begin{bmatrix} 
        \rho \mu_{{\rm nl},i} \\ 
        \rho u_1 \mu_{{\rm nl},i} \\ 
        \rho u_2 \mu_{{\rm nl},i} \\ 
        \rho u_3 \mu_{{\rm nl},i}
    \end{bmatrix}
        . 
    \label{eqn:PFDC_Fluxes}
\end{equation}
The flux $\bm{F}_i^{{\rm DC},u}$ is based on the deviatoric strain rate tensor of compressible fluids, as proposed in~\cite{Guermond2011-fp}. It remains active in both the bulk and interfacial regions to suppress instabilities arising from sharp velocity gradients. However, prior studies have shown that $\bm{F}_i^{{\rm DC},u}$ alone is insufficient to stabilize the solution; additional regularization based on the gradients of density or thermodynamic quantities is needed \cite{Guermond2008-bs}. Inspired by the Brenner regularization of the Navier–Stokes equations~\cite{Brenner2005-fy}, we include $\bm{F}_i^{{\rm DC},\rho}$ in the bulk fluid region. This flux has been shown to effectively suppress oscillations in the presence of sharp density gradients~\cite{Upperman2019-ed, Yamaleev2023-cw, Mukherjee2023-pv}. Since both $\bm{F}_i^{{\rm DC},u}$ and $\bm{F}_i^{{\rm DC},\rho}$ have been used in conventional compressible flow and share the same units of m$^2$/s as $\kappa_{\rm DC}$, we set
\begin{equation}
    \kappa_{\rm DC}^{u} = \kappa_{\rm DC}
    \label{eqn:kappa_DC_u}
\end{equation}
and 
\begin{equation}
    \kappa_{\rm DC}^{\rho} = \kappa_{\rm DC}.
    \label{eqn:kappa_DC_rho}
\end{equation}
When the solution enters the interfacial region, we replace $\bm{F}_i^{{\rm DC},\rho}$ by $\bm{F}_i^{{\rm DC},\mu}$, which is motivated by the concept of the non-local chemical potential defined in \cref{eqn:EulerLagrange}. The associated coefficient $\kappa_{\rm DC}^{\mu}$ has units of time and thus requires special consideration. Following the design philosophy of \cref{eqn:kappa_DC,eqn:kappa_DC_tilde,eqn:kappa_DC_hat}, we define
\begin{equation}
    \kappa_{\rm DC}^{\mu} = \min \left(\tilde{\kappa}_{\rm DC}^{\mu}, \hat{\kappa}_{\rm DC}^{\mu}\right),
    \label{eqn:kappa_DC_mu}
\end{equation} 
where 
\begin{equation}
    \tilde{\kappa}_{\rm DC}^{\mu} = C_{\rm DC}\left[\frac{{\bf Res}^{\intercal} \bm{\mathcal{N}} {\bf Res}}{G_{ij} \left(\bm{F}_i^{{\rm DC},\mu }\right)^{\intercal} \bm{\mathcal{N}} \left(\bm{F}_j^{{\rm DC},\mu }\right)}\right]^{\frac{1}{2}},
  \label{eqn:kappa_DC_mu_tilde}
\end{equation}
and
\begin{equation}
  \hat{\kappa}_{\rm DC}^{\mu} = \left[\lambda F \rho \bm{G} : \bm{G} + p_{,\rho}^{+} \mbox{tr}\left(\bm{G}\right) \right]^{-1/2}.
  \label{eqn:kappa_DC_mu_hat}
\end{equation}

\keyidea{Discuss the advantage of PFDC} If we substitute \cref{eqn:PFDC} into \cref{eqn:DVS_CompactForm_Mod} and follow the same procedure outlined in \cref{sec:EnergyDissipation}, we obtain the following modified free energy dissipation law in the interfacial region:
\begin{equation}
    \frac{d}{dt} \mathcal{E}(\rho, \rho \bm{u}) = - \int_\Omega \nabla \bm{u} : \overline{\mu}(\rho) \bm{D} {\rm d}\Omega - \int_\Omega \nabla \bm{u} : \eta \rho \kappa_{\rm DC}^{u} \bm{D} {\rm d}\Omega - \int_\Omega\eta \rho \kappa_{\rm DC}^{\mu} |\nabla\mu_{\rm nl}|^2 {\rm d}\Omega \leq 0.
    \label{eqn:EnergyDissipation_PFDC2}
\end{equation}
From \cref{eqn:EnergyDissipation_PFDC2}, we observe that the proposed operator \cref{eqn:PFDC} guarantees satisfaction of the free energy dissipation law at liquid-vapor interfaces, thereby resolving the limitations associated with \cref{eqn:DC_classical,eqn:DC_classical_wScale}. Since all three components of \cref{eqn:PFDC} are scaled by $\eta$, the proposed operator vanishes in the incompressible limit, ensuring solution accuracy in the bulk liquid. Additionally, when the liquid-vapor interface reaches equilibrium, the non-local chemical potential $\mu_{\rm nl}$ becomes spatially uniform, as described in \cref{eqn:EulerLagrange}. Under this condition, the final term in \cref{eqn:EnergyDissipation_PFDC2} vanishes, rendering the operator less dissipative at equilibrium. Because \cref{eqn:PFDC} integrates the structure of the classical DC operator with the Euler-Lagrange condition from phase-field theory, we refer to this new design as the {\Methodfull} ({\MethodAbr}) operator.

\begin{remark}
    There are alternative designs of the PF/DC operator that also satisfy the free energy dissipation law in the interfacial region. However, in our study, we found that \cref{eqn:PFDC} produces the most stable results when the liquid-vapor interface is severely underresolved, a situation that commonly occurs in engineering applications. For further details, see \cref{apx:AlternativeDesign}.
\end{remark}

\section{Application of discontinuity capturing operator to direct van der Waals simulation}
\label{sec:PFDC_DVS}

Direct van der Waals Simulation (DVS)~\cite{Hu2025-by, Hu2023-zr} is a computational framework for studying phase-transforming fluids. It augments the Navier-Stokes-Korteweg (NSK) equations with the thickened interface method (TIM)~\cite{Nayigizente2021-yy} and a residual-based stabilization technique. In this section, we first present the weak formulation associated with the DC-augmented NSK equations, \cref{eqn:DVS_CompactForm_Mod}. We then introduce the dispersive-SUPG operator designed to stabilize numerical solutions that involve mixed hyperbolic-dispersive wave structures. Finally, we describe the complete numerical formulation and the solution procedure used in this work.

\subsection{Weak formulation}
Following \cite{Hu2025-by}, we define our variable of interest as $\bm{Y} = \left[\log(\rho), u_1, u_2, u_3\right]^\intercal$ (see Remark \ref{remark:variable}), and rewrite \cref{eqn:DVS_CompactForm_Mod} in a quasi-linear form:
\begin{equation}
    \bm{{A}}_0 \bm{Y}_{,t} + \bm{{A}}_i^{\rm {hype \textbackslash p}} \bm{Y}_{,i} + \bm{{A}}_i^{p} \bm{Y}_{,i} + \bm{{F}}^{\rm disp}= \left(\bm{{K}}_{ij} \bm{Y}_{,j}\right)_{,i} + \bm{{F}}^{\rm DC}_{i,i}.
    \label{eqn:DVS_QuasiLinear}
\end{equation}
The transformation matrices are defined as:
\begin{alignat}{2}
    &\bm{A}_0 = \frac{\partial \bm{{U}}}{\partial \bm{Y}} \mbox{, } \\
    &\bm{{A}}_i^{\rm hype} = \bm{{A}}_i^{{\rm {hype \textbackslash p}}} + \bm{{A}}_i^{p} =  \frac{\partial \bm{{F}}^{\rm {hype \textbackslash p}}_i}{\partial \bm{Y}} + \frac{\partial \bm{{F}}^{\rm p}_i}{\partial \bm{Y}}, \\
    &\bm{{K}}_{ij}{\bm{Y}_{,j}} = {\bm{{F}}_i^{\rm diff}}\mbox{, }\\
    &\bm{A}_i^{\rm disp}\bm{A}_0^{-1}\left(\bm{A}_0\bm{Y}_{,i}\right)_{,jj} = \bm{F}^{\rm disp}.
\end{alignat}
For explicit expressions of these matrices, we refer the reader to Appendix C of \cite{Hu2025-by}. Direct numerical discretization of \cref{eqn:DVS_QuasiLinear} would require globally higher-order continuous basis functions. To avoid this constraint and enable the use of standard $C^0$-continuous finite elements, we introduce an auxiliary (split) variable and rewrite \cref{eqn:DVS_QuasiLinear} as:
\begin{alignat}{2}
        &\bm{{A}}_0 \bm{Y}_{,t} + \bm{{A}}_i^{\rm {hype \textbackslash p}} \bm{Y}_{,i} + \bm{F}^{\mathcal{M}}= \left(\bm{{K}}_{ij} \bm{Y}_{,j}\right)_{,i} + \bm{{F}}^{\rm DC}_{i,i},\label{eqn:DVS_QuasiLinear_Split} \\
        &\mathcal{M} = \mu_{\rm nl} - \frac{|\bm{u}|^2}{2},
\end{alignat}
where the newly introduced flux is defined as 
\begin{equation}
    \bm{F}^{\mathcal{M}} = 
    \begin{bmatrix} 
        0 \\ 
        \rho \mathcal{M}_{,1} + \rho u_{j}u_{j,1} \\ 
        \rho \mathcal{M}_{,2} + \rho u_{j}u_{j,2} \\ 
        \rho \mathcal{M}_{,3} + \rho u_{j}u_{j,3}
      \end{bmatrix}
      .
\end{equation}
Using the identity in \cref{eqn:EnergyDissipation_Property2} and basic manipulations, it can be shown that $\bm{F}^{\mathcal{M}} = \bm{A}_i^{p} \bm{Y}_{,i} + \bm{F}^{\rm disp}$. With the split variable formulation, the {\MethodAbr} flux associated with $\mu_{\rm nl}$ in \cref{eqn:PFDC_Fluxes} becomes:
\begin{equation}
    \bm{F}_i^{{\rm DC},\mu} =  \begin{bmatrix} 
        \rho \left(\mathcal{M}_{,i} + u_{j}u_{j,i}\right)\\ 
        \rho u_1 \left(\mathcal{M}_{,i} + u_{j}u_{j,i}\right)\\ 
        \rho u_2 \left(\mathcal{M}_{,i} + u_{j}u_{j,i}\right)\\ 
        \rho u_3 \left(\mathcal{M}_{,i} + u_{j}u_{j,i}\right)
    \end{bmatrix}
    .
    \label{eqn:PFDC_Flux_mod}
\end{equation}
The strong-form of the residual used to evaluate the dissipation coefficients in \cref{eqn:kappa_DC_tilde,eqn:kappa_DC_mu_tilde} is modified as follows:
\begin{equation}
\label{eqn:residual}
{\bf Res}(\bm{Y},\mathcal{M}) = \bm{{A}}_0 \bm{Y}_{,t} + \bm{{A}}_i^{\rm {hype \textbackslash p}} \bm{Y}_{,i} + \bm{F}^{\mathcal{M}} - \left(\bm{{K}}_{ij} \bm{Y}_{,j}\right)_{,i}.
\end{equation}
We assume the computational domain $\Omega$ is partitioned into $N_{\rm el}$ elements, each denoted by $\Omega^e$. The Galerkin operator of \cref{eqn:DVS_QuasiLinear_Split} is given by
\begin{alignat}{2}
    \bm{B}_{\rm Galerkin}\left(\{\bm{W},Q\},\{\bm{Y},\mathcal{M}\}\right) &= \int _\Omega \bm{W} \cdot \left(\bm{{A}}_0 \bm{Y}_{,t} + \bm{{A}}_i^{\rm {hype \textbackslash p}} \bm{Y}_{,i} + \bm{F}^{\mathcal{M}}\right) {\rm d} \Omega + \int_\Omega Q\left(\mathcal{M} - \mu + \frac{|\bm{u}|^2}{2}\right) {\rm d}\Omega \nonumber\\    
    &+ \int_\Omega \bm{W}_{,i} \cdot \bm{{F}}_i^{\rm diff} {\rm d} \Omega + \sum ^ {N_{\rm el}}_{e=1} \int _{\Omega^e}  \bm{W}^h_{,i} \cdot \bm{F}^{\rm DC}_{i} {\rm d}\Omega^e - \int_\Omega \lambda F Q_{,i} \rho_{,i}{\rm d}\Omega \nonumber\\    
    &- \int _\Gamma \bm{W} \cdot \bm{{F}}_i^{\rm diff} n_i {\rm d} \Gamma + \int _\Gamma \lambda F Q \rho_{,i}n_i {\rm d} \Gamma.
    \label{eqn:WeakForm}
\end{alignat}
Here, the weight functions $\{\bm{W}, Q\}$ belong to a suitable test space, $\Gamma$ denotes the boundary of $\Omega$, and $n_i$ is the $i$th component of the unit outward normal to $\Gamma$.

\begin{remark}
    Across the liquid–vapor interface, fluid density can change over 5 orders of magnitude over a small spatial region. Even a small numerical oscillation can potentially drive density to a negative value and lead to a non-physical solution. To avoid this issue, we use $\log\left(\rho\right)$ instead of $\rho$ for the variables of interest $\bm{Y}$. This choice ensures the numerical solution of density remains strictly positive throughout the entire simulation.
    \label{remark:variable}
\end{remark}

\subsection{Dispersive-SUPG operator}
In addition to sharp layers, the solution of \cref{eqn:DVS_CompactForm} can also exhibit mixed hyperbolic/dispersive wave structures. When these structures dominate the flow physics, the standard Galerkin method can produce spurious nodal oscillations, which may grow over time and ultimately cause the numerical simulation to diverge. To address this issue, we augment \cref{eqn:WeakForm} with a dispersive-SUPG (D-SUPG) operator~\cite{Hu2025-by}, a modification of the classical SUPG stabilization~\cite{Brooks1982-og, Shakib1991-ji}. This method is capable of producing high-order accurate results when the solution is dominated by either hyperbolic or dispersive waves. The D-SUPG operator is defined as: 
\begin{equation}
    \bm{B}_{\rm D-SUPG}\left(\bm{W},\{\bm{Y},\mathcal{M}\}\right) = \sum ^ {N_{\rm el}} _ {e=1} \int _{\Omega^e} \left(\bm{A}_i^{*T} \bm{W}_{,i}\right) \cdot \left(\bm{{A}}_0 ^{-1}\bm{\tau}_{\rm Stab}\right) {\bf Res}(\bm{Y},\mathcal{M}) {\rm d} \Omega^e.
      \label{eqn:DVS_SUPG}
\end{equation}
where the stabilization matrix is given by
\begin{equation}
  \bm{{\tau}}_{\rm Stab} = \left[\frac{4\bm{I}}{\Delta t^2}  + G_{ij}\left(\bm{{A}}^{*}_i \bm{{A}}_0^{-1}\right) \left(\bm{{A}}^{*}_j \bm{{A}}_0^{-1}\right) + C_I G_{ij} G_{kl} \left(\bm{{K}}_{ik} \bm{{A}}_0^{-1}\right)\left(\bm{{K}}_{jl} \bm{{A}}_0^{-1}\right)\right].
  \label{eqn:tau_SUPG}
\end{equation}
Here, $\Delta t$ is the time-step size and $C_I$ is a positive constant obtained from an element-wise inverse estimate \cite{Johnson2012-ms}. The matrices $\bm{{A}}^{*}_i $ defined by
\begin{equation}
    \bm{{A}}^{*}_i =  \left(\bm{I} - \bm{C}^{d/h}_i\right) \bm{{A}}^{\rm hype}_i = \bm{{A}}^{\rm hype}_i - G_{ii}\bm{A}_i^{\rm disp} \quad \mbox{(no sum on $i$)},
\end{equation}
quantify the relative strength of dispersive waves compared to hyperbolic waves and dynamically adjust the directional bias toward the upwind direction of the dominant wave type to enhance numerical stability. The matrix inverse square root in \cref{eqn:tau_SUPG} can be approximated numerically using Denman-Beavers iteration \cite{Denman1976-nm, Xu2017-gq}.

\subsection{Complete formulation and solution procedure}
\label{sec:num_procedure}
The proposed algorithm for DVS is constructed by combining \cref{eqn:WeakForm,eqn:DVS_SUPG}. We use a finite element space $\mathcal{V}^h$ that satisfies the Dirichlet boundary conditions and an analogous discrete space $\mathcal{V}_0^h$ that satisfies homogeneous conditions at the Dirichlet boundary. Spatial discretization is performed using standard $C^0$-continuous finite elements. The semi-discretized problem is defined as follows: find $\{\bm{Y}^h,\mathcal{M}^h\} \in \mathcal{V}^h$, such that for all $\{\bm{W}^h,Q^h\} \in \mathcal{V}^h_0$ 
\begin{alignat}{2}
    \bm{B}_{\rm Galerkin}\left(\{\bm{W},Q\},\{\bm{Y},\mathcal{M}\}\right) &= \int _\Omega \bm{W} \cdot \left(\bm{{A}}_0 \bm{Y}_{,t} + \bm{{A}}_i^{\rm {hype \textbackslash p}} \bm{Y}_{,i} + \bm{F}^{\mathcal{M}}\right) {\rm d} \Omega + \int_\Omega Q\left(\mathcal{M} - \mu + \frac{|\bm{u}|^2}{2}\right) {\rm d}\Omega \nonumber\\    
    &+ \int_\Omega \bm{W}_{,i} \cdot \bm{{F}}_i^{\rm diff} {\rm d} \Omega + \sum ^ {N_{\rm el}}_{e=1} \int _{\Omega^e}  \bm{W}^h_{,i} \cdot \bm{F}^{\rm DC}_{i} {\rm d}\Omega^e - \int_\Omega \lambda F Q_{,i} \rho_{,i}{\rm d}\Omega \nonumber\\    
    &+\sum ^ {N_{\rm el}} _ {e=1} \int _{\Omega^e} \left(\bm{A}_i^{*T} \bm{W}_{,i}\right) \cdot \left(\bm{{A}}_0 ^{-1}\bm{\tau}_{\rm Stab}\right) {\bf Res}(\bm{Y},\mathcal{M}) {\rm d} \Omega^e \nonumber \\
    &- \int _\Gamma \bm{W} \cdot \bm{{F}}_i^{\rm diff} n_i {\rm d} \Gamma + \int _\Gamma \lambda F Q \rho_{,i}n_i {\rm d} \Gamma.
    \label{eqn:DVS_Formulation}
\end{alignat}
Temporal discretization is performed using the generalized-$\alpha$ method~\cite{Jansen2000-gg} with $\varrho_\infty = 0.5$. At each time step, the resulting nonlinear system is solved using a Newton-Raphson algorithm with line search. Nonlinear convergence is assessed at every time step using a relative residual tolerance of $10^{-4}$ for $\bm{Y}^h$ and a relative $L^2$-norm change tolerance of $10^{-6}$ for $\mathcal{M}^h$. The linearized systems are solved using the Generalized Minimal Residual (GMRES) method~\cite{Saad1986-ap}, preconditioned with an additive Schwarz method~\cite{Cai1999-up}. Each block subdomain employs an incomplete LU (ILU) factorization~\cite{Dupont1968-al} with a level-2 fill for sparsity pattern. The Eisenstat-Walker method~\cite{Eisenstat1996-jd} is used to dynamically compute the relative tolerance for the linear solver. All implementations are performed in the open-source libraries PETSc~\cite{Balay2021-vc} and PetIGA~\cite{Dalcin2013-dl}.

Since the time integration scheme is implicit, and our experience suggests that the primary constraints on $\Delta t$ arise from the accuracy and solvability of the nonlinear algebraic system, we employ an adaptive time-stepping strategy to maintain efficient Newton solver performance. Specifically, if the Newton solver converges within three iterations, $\Delta t$ is increased by 5\%. If convergence requires five or more iterations, $\Delta t$ is decreased by 5\%. In the event of divergence, $\Delta t$ is reduced by 75\% and we restart the time-stepping from the diverged time step.

\section{Numerical results}
\label{sec:Results}

In this section, we conduct a series of numerical studies to evaluate the performance of \cref{eqn:DVS_Formulation} under various designs of the DC operator. When the flux $\bm{F}^{\rm DC}_i$ in \cref{eqn:DVS_Formulation} is set to \cref{eqn:DC_classical,eqn:DC_classical_wScale}, we refer to the resulting methods as DC (no scale) and DC (with scale), respectively. When $\bm{F}^{\rm DC}_i$ is defined by the proposed formulation in \cref{eqn:PFDC,eqn:PFDC_Fluxes,eqn:PFDC_Flux_mod}, we denote the method as {\Methodfull} ({\MethodAbr}). We first examine the accuracy and stability of the three methods in a one-dimensional domain. Our results demonstrate that only {\MethodAbr} achieves the optimal convergence rate while faithfully preserving the flow physics in both the interfacial and bulk regions. This finding is further supported by a two-dimensional simulation involving interacting non-parallel waves and a dynamically evolving liquid-vapor interface. Finally, we apply DVS to simulate cavitating flow over a three-dimensional bluff body, and we compare the numerical results against experimental data. Relative to the classical DC operators, {\MethodAbr} significantly enhances the predictive capability of DVS for engineering applications. Throughout this study, we fix $\lambda = 5.0094\times 10^{-17}$ m$^7$/kg/s$^2$, which yields accurate predictions of water surface tension across a range of temperatures.

\subsection{Accuracy study}

\begin{figure}[htbp]
    \centering
    \includegraphics[keepaspectratio=true,width=\linewidth]{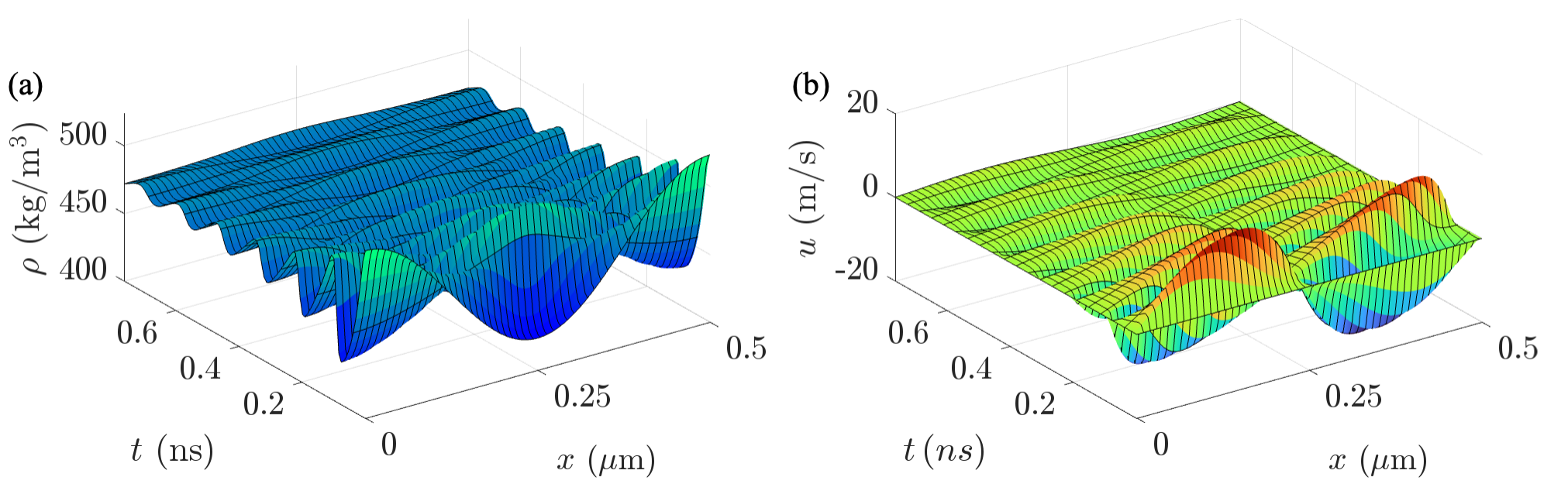}
    \caption{Time evolution of (a) density and (b) velocity of the reference solution with $\rho^{\rm ref} = \rho^s_l + 50$ kg/m$^3$ on a one-dimensional periodic domain. Given the initial density distribution, the fluid remains entirely in the liquid phase, and the perturbation is insufficient to induce phase transformation. As a result, the solution remains single-phase and exhibits sinusoidal oscillations whose amplitude gradually decreases over time due to viscous dissipation.}
    \label{fig:Convergence_Data_BulkLiquid}
\end{figure}

\begin{figure}[htbp]
    \centering
    \includegraphics[keepaspectratio=true,width=\linewidth]{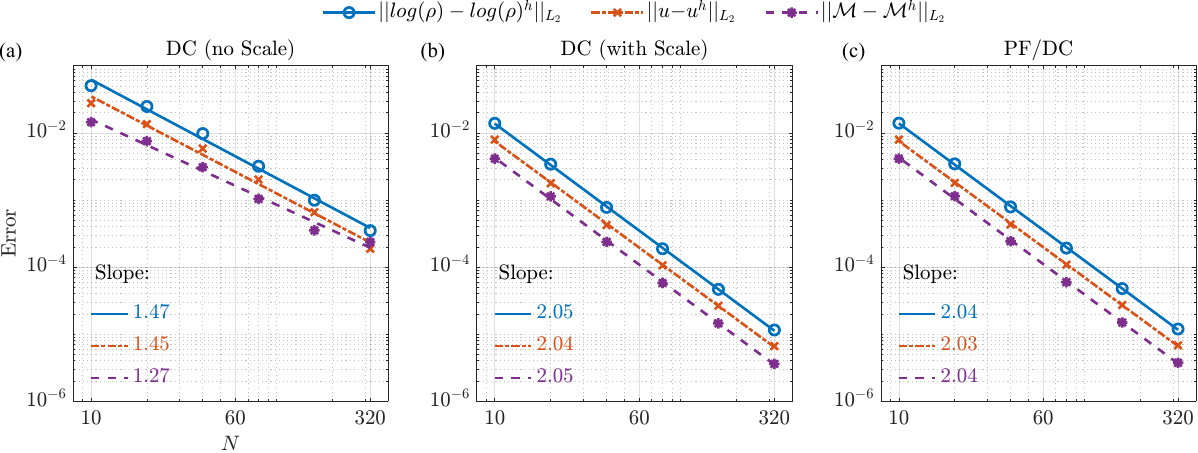}
    \caption{$L_2$ norm of the error in the numerical solution at $t = 0.076$ ns with $\rho^{\rm ref} = \rho^s_l + 50$ kg/m$^3$, evaluated against the reference solution for different DC operator choices in \cref{eqn:DVS_Formulation}. The DC (no scale) operator yields a suboptimal convergence rate, while both the DC (with scale) and {\MethodAbr} operators achieve optimal rate. These results provide numerical evidence supporting the effectiveness of scaling the DC operator based on fluid compressibility.}
    \label{fig:Convergence_Error_BulkLiquid}
\end{figure}

Here, we study the accuracy of various DC operators when the solution is sufficiently smooth. A convergence analysis is performed in a one-dimensional domain of length $L_0 = 0.5$ $\mu$m with periodic boundary conditions. We temporarily modify the viscosity to satisfy the viscosity-capillarity criterion \cite{Hagan1983-hi, Slemrod1983-ro}, using
\begin{equation}
    \overline{\mu} = \rho \sqrt{4\lambda F \rho}.
    \label{eqn:visco_capillarity}
\end{equation} 
The domain is uniformly discretized using $N = 10, 20, 40, 80, 160$ and $320$ linear elements. Since an exact solution is not available, the numerical solution obtained on a fine mesh with $N = 1280$ linear elements is used as the reference. As shown in \cite{Hu2025-by}, \cref{eqn:DVS_Formulation} without any DC operator yields the optimal convergence rate in both the bulk and interfacial regions; thus, it is employed to compute the reference solution. The convergence study is conducted over the time interval $t = 0$ ns to $t = 0.764$ ns, using a fixed small time step of $\Delta t = 1.53 \times 10^{-8}$ ns to minimize temporal discretization error. The fluid temperature is set to $T = 645$ K with $F = 10$, corresponding to saturation liquid and vapor densities of $423.7$ kg/m$^3$ and $224.3$ kg/m$^3$, respectively. The initial velocity is set to zero, and the initial density profile is prescribed as
\begin{equation}
\rho(x,0) = \rho^{\rm ref} + 50 \cos\left(\frac{2\pi x}{L_0}\right) \ \text{kg/m}^3,
\end{equation}
where $\rho^{\rm ref}$ is a constant base density value. We consider two values of $\rho^{\rm ref}$, each representing scenarios that do not arise in conventional compressible flow but are commonly encountered in solutions of \cref{eqn:DVS_CompactForm}. First, we set $\rho^{\rm ref} = \rho^s_l + 50$ kg/m$^3$, such that the entire fluid is initially in the bulk liquid state and nearly incompressible. In \cref{fig:Convergence_Data_BulkLiquid}, we plot the reference (a) density and (b) velocity fields as functions of spatial coordinate $x$ and time $t$. Because the initial perturbation is insufficient to trigger phase transformation, the system remains in the liquid state and exhibits sinusoidal oscillations whose amplitude gradually decays over time due to viscous dissipation. \Cref{fig:Convergence_Error_BulkLiquid} presents the $L_2$ norm of the error in $\log(\rho)$, ${u}$, and $\mathcal{M}$ at $t = 0.076$ ns, evaluated with respect to the reference solution for different choices of DC operator. Although the DC (no scale) operator converges due to the consistency of the formulation, it yields a suboptimal convergence rate between 1.27 and 1.47. This loss of accuracy is attributed to the fact that \cref{eqn:DC_classical} does not vanish in the incompressible limit. In contrast, both the DC (with scale) and {\MethodAbr} operators achieve optimal convergence rates, demonstrating the necessity and effectiveness of scaling based on fluid compressibility.

\begin{figure}[htbp]
    \centering
    \includegraphics[keepaspectratio=true,width=\linewidth]{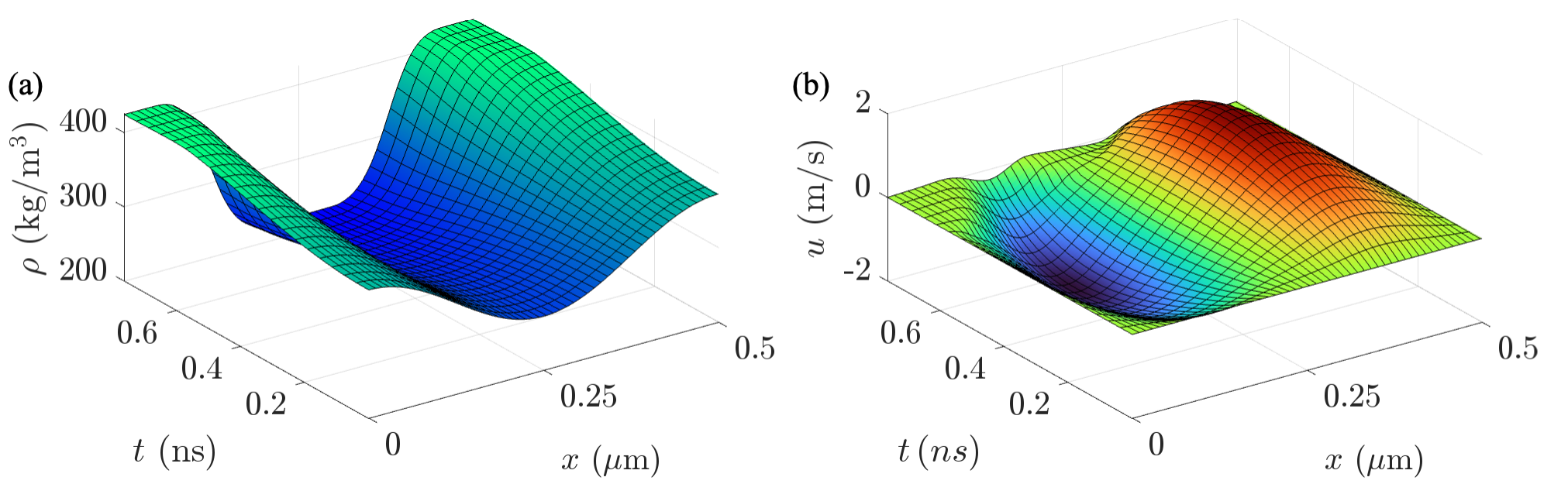}
    \caption{Time evolution of (a) density and (b) velocity of the reference solution with $\rho^{\rm ref} = 322.6$ kg/m$^3$ on a one-dimensional periodic domain. The fluid is initially in the interfacial regime, allowing perturbations to grow as time progresses. Once sufficiently amplified, the perturbation leads to the formation of distinct liquid and vapor phases, with stable interfaces developing in the domain.}
    \label{fig:Convergence_Data_Interface}
\end{figure}

\begin{figure}[htbp]
    \centering
    \includegraphics[keepaspectratio=true,width=\linewidth]{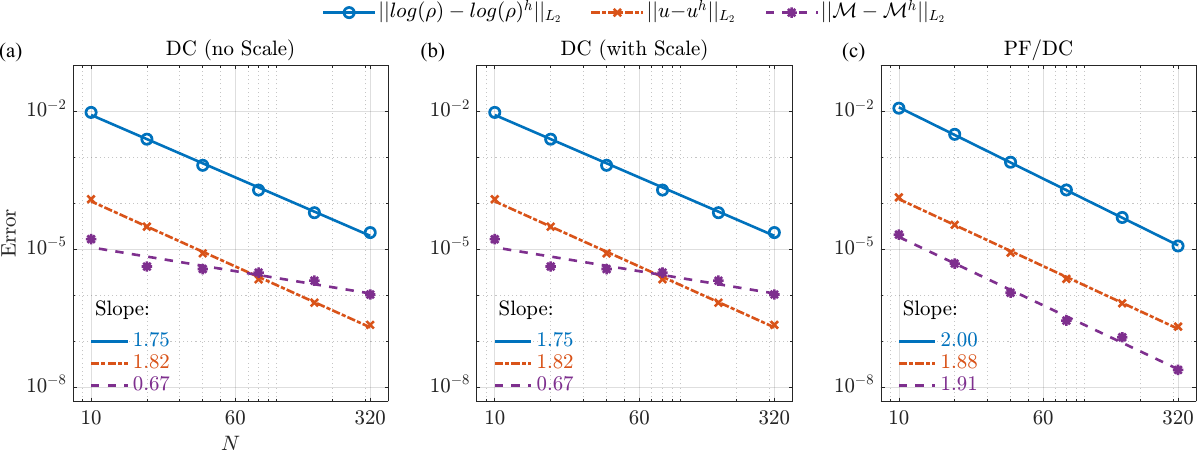}
    \caption{$L_2$ norm of the error in the numerical solution at $t = 0.076$ ns with $\rho^{\rm ref} = 322.6$ kg/m$^3$, evaluated against the reference solution for different DC operator choices in \cref{eqn:DVS_Formulation}. Both the DC (no scale) and DC (with scale) operators exhibit suboptimal convergence rates, whereas the {\MethodAbr} operator achieves optimal convergence rate for $\log{(\rho)}$ and a near-optimal convergence rate for $u$ and $\mathcal{M}$. These results highlight the importance of satisfying free energy stability in the development of accurate numerical methods.}
    \label{fig:Convergence_Error_Interface}
\end{figure}

We next evaluate the performance of various DC operators during phase transformation. We set $\rho^{\rm ref} = 322.6$ kg/m$^3$, placing the initial fluid state within the interfacial regime. In \cref{fig:Convergence_Data_Interface}, we plot the reference (a) density and (b) velocity. Unlike the behavior observed in \cref{fig:Convergence_Data_BulkLiquid}, the initial sinusoidal perturbation grows over time due to the negative value of $\frac{d p}{d \rho}$. As the fluid expands, a nucleation site forms at the center of the domain, ultimately leading to the development of a stable vapor slug within the region $x \in [0.125, 0.375]$ $\mu$m. In \cref{fig:Convergence_Error_Interface}, we present the convergence behavior of the different DC operators at $t = 0.076$ ns. Both the DC (no scale) and DC (with scale) operators yield converged solutions but exhibit suboptimal rates between 0.67 and 1,82. We attribute this reduced accuracy to the lack of free energy stability in \cref{eqn:DC_classical,eqn:DC_classical_wScale}. In contrast, {\MethodAbr} attains the optimal convergence rate for $\log{(\rho)}$ and a near-optimal rate for $u$ and $\mathcal{M}$, demonstrating its accuracy in both the bulk liquid and interfacial regions.

\subsection{Stability study}
\label{sec:StabilityStudy}

\begin{figure}[htbp]
    \centering
    \includegraphics[keepaspectratio=true,width=0.75\linewidth]{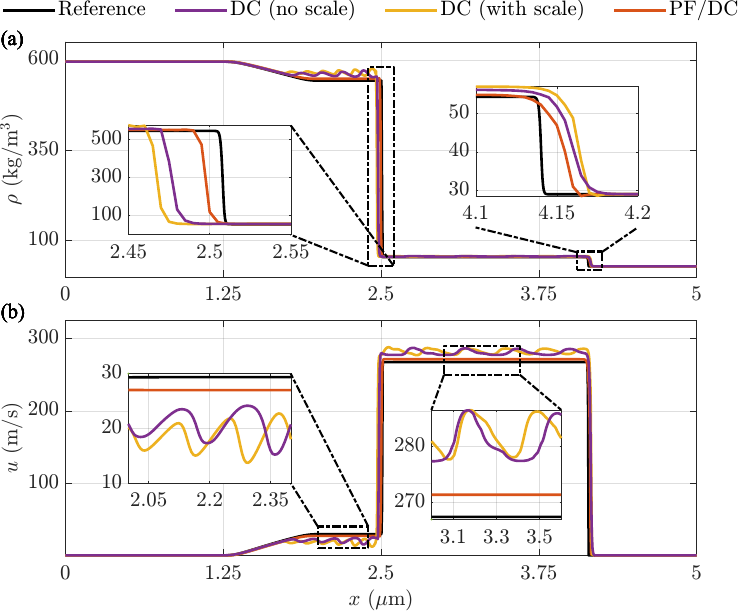}
    \caption{Numerical solutions of (a) density and (b) velocity for a one-dimensional multiphase wave propagation problem using various DC operators on a coarse mesh with $N = 10^3$, compared against the reference solution. The {\MethodAbr} operator yields the most accurate prediction of the liquid-vapor interface location and wave propagation speed in the bulk vapor. Moreover, {\MethodAbr} correctly captures the overall flow physics, whereas the DC (no scale) and DC (with scale) operators generate unphysical oscillations due to a lack of free energy stability.}
    \label{fig:MultiphaseTube_N1000}
\end{figure}

\begin{figure}[htbp]
    \centering
    \includegraphics[keepaspectratio=true,width=0.75\linewidth]{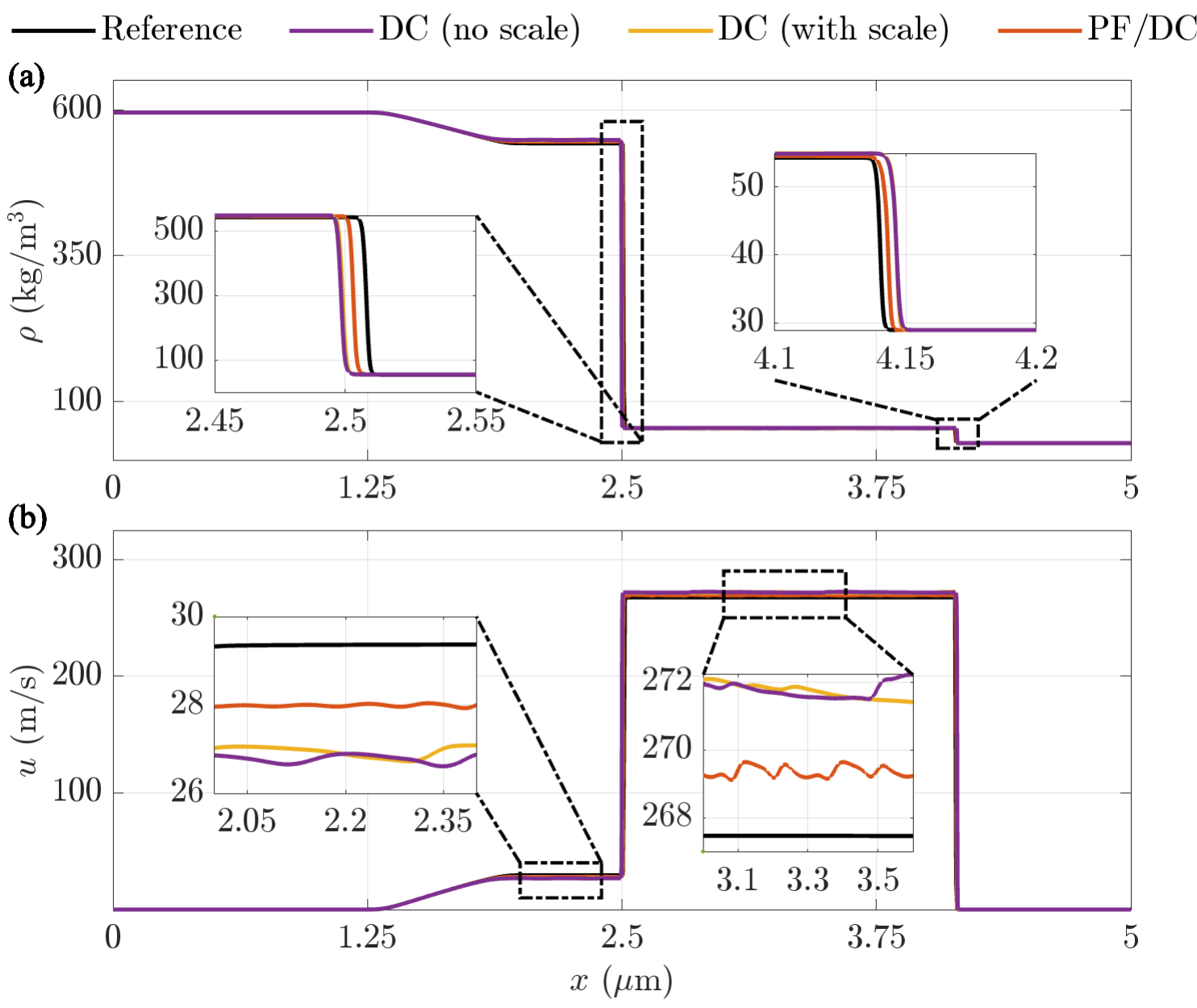}
    \caption{Numerical solutions of (a) density and (b) velocity for a one-dimensional multiphase wave propagation problem using various DC operators on a fine mesh with $N = 10^4$, compared against the reference solution. While all DC operators capture the correct qualitative flow behavior, {\MethodAbr} yields the most accurate prediction of both the liquid–vapor interface and the wavefront locations. These results provide additional numerical evidence supporting the accuracy of the {\MethodAbr} formulation.}
    \label{fig:MultiphaseTube_N10000}
\end{figure}

Here, we investigate the stability of numerical solutions initialized with overcompressed liquid and overexpanded vapor states. Simulations are performed in a one-dimensional domain of length $L_0 = 5$ $\mu$m, uniformly discretized using linear elements. We impose no-flow boundary conditions by setting $u = 0$ and $\nabla \rho \cdot \bm{n} = 0$ at all boundaries. The temperature is fixed at $T = 625$ K with $F = 1$, corresponding to saturation liquid and vapor densities of $567.1$ kg/m$^3$ and $115.7$ kg/m$^3$, respectively. We initialize the system using a smoothed profile commonly adopted in one-dimensional shock tube problems, with $u(x, 0) = 0$ m/s and
\begin{equation}
    \rho\left(x,0\right) = \frac{\rho^{\rm left} + \rho^{\rm right}}{2.0} + \frac{\rho^{\rm left} - \rho^{\rm right}}{2.0} \tanh\left(\frac{x-0.5}{0.5\times 10^{-4} L_0}\right)  \ \text{kg/m}^3,
\end{equation}
where $\rho^{\rm left} = 1.05 \rho_l^s$ and $\rho^{\rm right} = 0.25 \rho_v^s$. Unlike classical Riemann problems, the presence of both hyperbolic and dispersive waves in this setting prevents analytical determination of wave evolution. To address this, we temporarily modify the viscosity using \cref{eqn:visco_capillarity} which ensures the existence of monotonic wave fronts. A reference solution is then computed using \cref{eqn:DVS_Formulation} without any DC operator on a highly resolved mesh with $N = 10^5$ linear elements. The resulting solution is free of spurious oscillations in both the density and velocity fields, and the wave structures remain fixed under further refinement. In \cref{fig:MultiphaseTube_N1000,fig:MultiphaseTube_N10000}, we plot the (a) density and (b) velocity profiles of the reference solution, alongside results obtained using various DC operators at $t = 2.865$ ns. Three distinct wave features are observed. In the region $x \in [1.25, 2.0]$ $\mu$m, an expansion fan forms in the bulk liquid. Near $x = 2.5$ $\mu$m, a stable liquid-vapor interface forms, across which the density changes, while the pressure remains the same on both sides of the interface. Finally, around $x = 4.15$ $\mu$m, a sharp wave front develops in the bulk vapor region, exhibiting characteristics similar to a shock wave in the classical Riemann problem.

To assess the robustness of the numerical method, we consider two different mesh resolutions. In \cref{fig:MultiphaseTube_N1000}, we present the solution obtained using $N = 10^3$ linear elements, where the liquid–vapor interface is severely under-resolved. Both the DC (no scale) and DC (with scale) operators yield inaccurate predictions of the liquid–vapor interface location and overestimate the propagation speed of the wave front in the bulk vapor region. More critically, due to the lack of free energy stability, both DC operators introduce unphysical oscillations at the liquid–vapor interface, which subsequently propagate into the surrounding fluid. The emergence of such spurious wave structures significantly compromises the reliability of the solution and may lead to numerical instability. In contrast, {\MethodAbr} predicts more accurate locations of both the interface and the wave front in the vapor phase. Moreover, it does not produce any unphysical flow features, highlighting the importance of free energy stability in maintaining the accuracy and robustness of the numerical method.
We then refine the mesh to $N = 10^4$ linear elements, ensuring that the liquid-vapor interface spans at least five grid points. The resulting solution is shown in \cref{fig:MultiphaseTube_N10000}. Due to variational consistency, all DC operators produce monotonic density fields as the strong-form residual decreases, with {\MethodAbr} yielding the most accurate prediction of both the liquid–vapor interface location and the wave propagation speed in the bulk vapor.

\subsection{Periodic oscillation of liquid-vapor interface}
\begin{figure}[htbp]
    \centering
    \includegraphics[keepaspectratio=true,width=\linewidth]{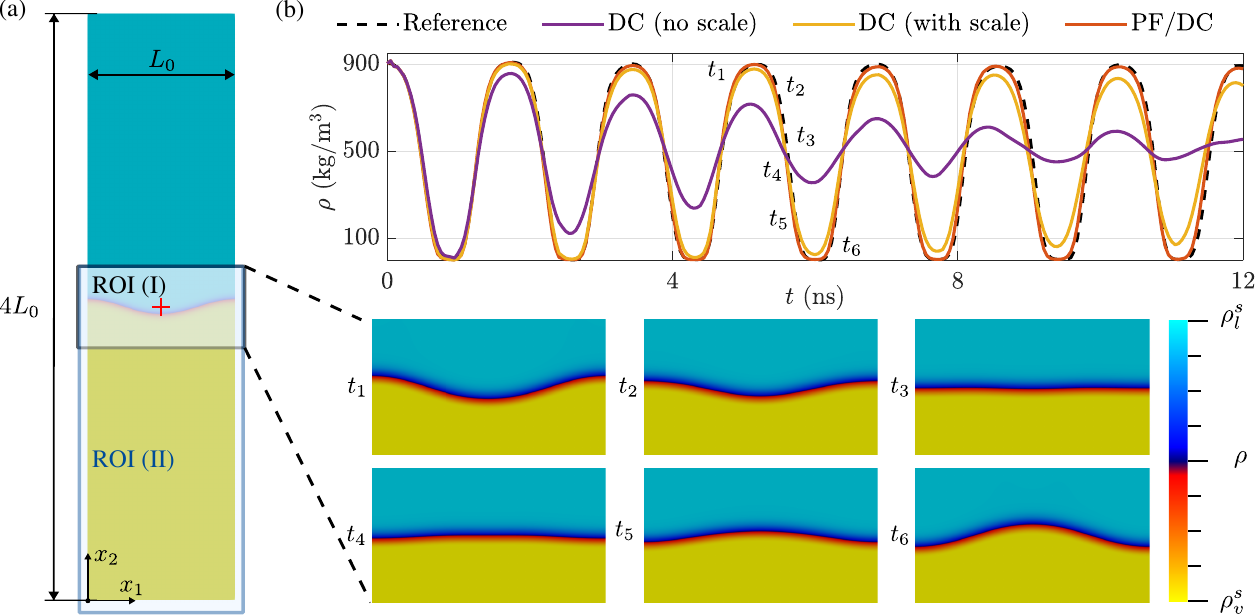}
    \caption{(a) Computational domain with two regions of interest (ROI): (I) the liquid–vapor interface and (II) the bulk vapor region. (b) Time evolution of the density at the center of the domain for various DC operators, along with snapshots of the reference density contours in ROI (I) from $t_1$ to $t_6$, covering half an oscillation cycle. The results show that all DC operators accurately capture the oscillation frequency, but only {\MethodAbr} maintains the oscillation amplitude over time.}
    \label{fig:Interface_Density}
\end{figure}

We now consider a two-dimensional domain to evaluate the performance of various DC operators in scenarios where non-parallel waves continuously interact with a dynamically evolving liquid–vapor interface. The computational domain is described in \cref{fig:Interface_Density}(a) where $L_0 = 0.1$ $\mu$m. The domain is uniformly discretized using $160 \times 640$ bilinear elements. We enforce $\bm{u} \cdot \bm{n} = 0$ and $\nabla \rho \cdot \bm{n} = 0$ at all boundaries. The fluid parameters are set to $F = 10$ and $T = 575$ K, corresponding to saturation densities of $\rho_l = 937.0$ kg/m$^3$ and $\rho_v = 1.4$ kg/m$^3$ for the liquid and vapor phases, respectively. To isolate the effect of numerical dissipation, we set $\overline{\mu} = 0$ to eliminate any physical damping. The initial velocity field is set to zero, and the initial density field is
\begin{equation}
    \rho_0\left(\bm{x},0\right) = \frac{\rho_l + \rho_v}{2} + \frac{\rho_l - \rho_v}{2} \tanh \left(0.01125 \frac{x_2 - x_{2,c}(x_1)}{\rho_c \sqrt{\lambda / P_c}}\right)  \ \text{kg/m}^3,
    \label{eqn:InterfaceProfile}
\end{equation}
where $P_c = 22.064$ MPa is the critical pressure. The center of the initial interface along the $x_2$ direction is given by
\begin{equation}
    x_{2,c}(x_1) = 2L_0 + 0.05L_0 \frac{1 - \cos\left(2\pi {x_1} /L_0\right)}{2}.
\end{equation}
Having established the stability and accuracy of {\MethodAbr}, we solve \cref{eqn:DVS_Formulation} with the flux definition given in \cref{eqn:PFDC,eqn:PFDC_Fluxes,eqn:PFDC_Flux_mod} on a refined mesh of $320 \times 1280$ bilinear elements to obtain a reference solution. In \cref{fig:Interface_Density}(b), we plot the time evolution of the fluid density at the center of the domain and include snapshots of the density contours at six time stamps spanning half an oscillation cycle. At $t = 0$ ns, because of the initial condition defined in \cref{eqn:InterfaceProfile} is out of equilibrium, the density field undergoes a brief period of localized oscillations as it relaxes toward an interface profile with a thickness and shape consistent with the governing equations. After this initial adjustment, the interface exhibits sustained periodic oscillations, with a frequency governed by the domain size, saturation densities, and surface tension, as described in \cite{Fyfe1988-jl}. In the absence of viscosity, \cref{eqn:EnergyDissipation} indicates that the total free energy of the system remains constant in time, and therefore the oscillation amplitude should remain unchanged across cycles. Accurately capturing this long-time oscillatory behavior is crucial for many engineering applications \cite{Ashgriz1990-ok}. Under the same mesh resolution, we find that all DC operators correctly reproduce the oscillation frequency. However, the DC (no scale) operator induces significant amplitude damping over successive cycles, indicating excessive numerical dissipation. The DC (with scale) operator provides significant improvement over DC (no scale), but still exhibits noticeable amplitude decay over time. In contrast, {\MethodAbr} produces a solution nearly indistinguishable from the reference, with virtually no decay in oscillation amplitude, highlighting its superior accuracy and minimal numerical dissipation.

\begin{figure}[htbp]
    \centering
    \includegraphics[keepaspectratio=true,width=\linewidth]{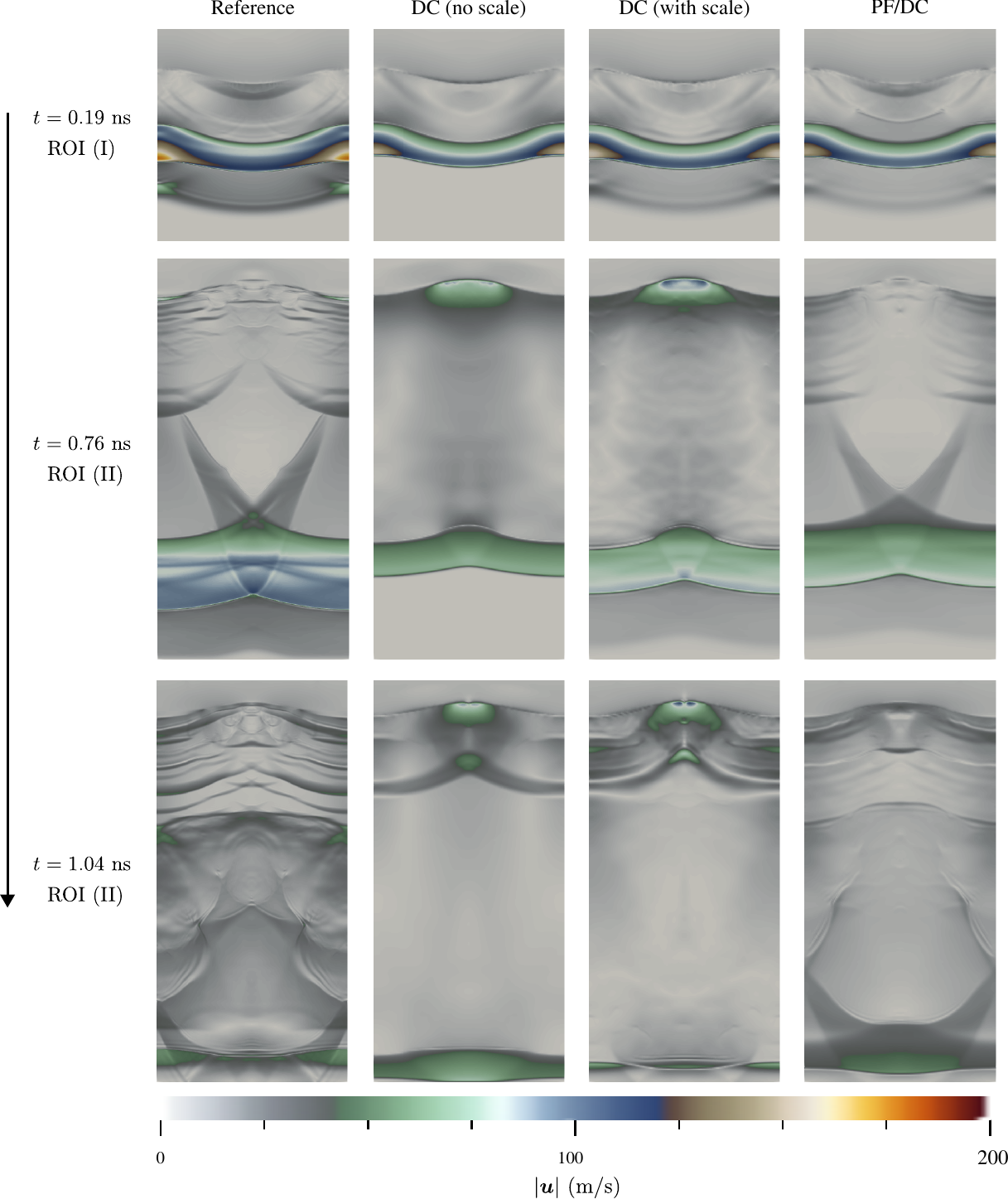}
    \caption{Snapshots of velocity contours in ROI (I) and (II) at $t = 0.19$ ns (first row), $t = 0.76$ ns (second row), and $t = 1.04$ ns (third row). The DC (no scale) operator exhibits excessive dissipation of high-frequency wave modes, resulting in rapid decay of oscillation amplitude. Furthermore, both DC (no scale) and DC (with scale) operators miss many wave structures and generate unphysical jets near the interface that are absent in the reference solution. In contrast, {\MethodAbr} preserves most wave structures and accurately captures the underlying physics.}
    \label{fig:Interface_Velocity}
\end{figure}

The performance difference among various DC operators becomes evident immediately after the simulation begins. In \cref{fig:Interface_Velocity}, we present three snapshots of the fluid velocity contours near $t = 0$ ns, focused on ROI (I) and (II) as defined in \cref{fig:Interface_Velocity}(a). At $t = 0.19$ ns, the solution begins to adjust the initial condition to recover the correct liquid-vapor interface profile. During this process, the interface emits a series of oscillatory wave trains that propagate into the bulk liquid and vapor regions. These waves eventually reflect off the impermeable boundaries and return momentum to the interface, sustaining its oscillatory motion without amplitude decay. While both DC (with scale) and {\MethodAbr} successfully capture the full extent of the wave train, DC (no scale) fails to resolve the high-frequency components and instead produces a much shorter wave train due to excessive numerical dissipation. This explains the rapid amplitude decay observed in its solution. By $t = 0.76$ ns, the wave trains in the reference solution have propagated away from the interface and evolved into complex wave structures. However, many of these features are missing in the solutions obtained with both DC (no scale) and DC (with scale). More critically, due to the lack of free energy stability in the interfacial regions, both DC (no scale) and DC (with scale) generate an unphysical jet near the interface that is not present in the reference solution. At $t = 1.04$ ns, this jet evolves into a secondary wave that propagates away from the interface, similar to the behavior observed in \cref{sec:StabilityStudy}. These spurious waves are eventually damped by the DC operators as they enter the bulk fluid and thereby maintaining overall numerical stability. However, this artificial attenuation comes at the cost of excessive dissipation of inertia, resulting in a gradual decay of the oscillation amplitude. In contrast, {\MethodAbr} captures the majority of the wave structures in the reference solution and accurately represents the underlying physics. These results further demonstrate that {\MethodAbr} outperforms classical DC operators in both accuracy and physical fidelity.

\subsection{Cavitating flow over 3D bluff body}
\begin{figure}[htbp]
    \centering
    \includegraphics[keepaspectratio=true,width=\linewidth]{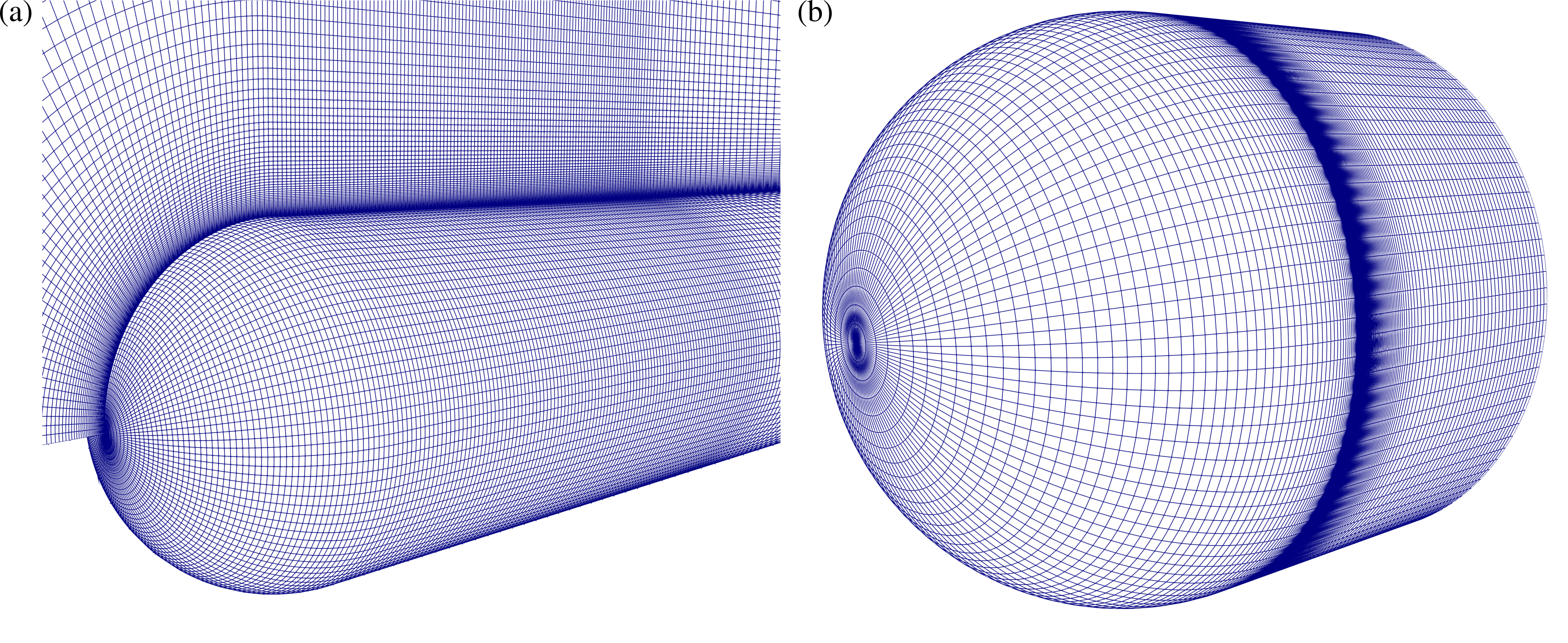}
    \caption{(a) Close-up view of the mesh near the hemispherical nose, including a cross-sectional slice normal to the wall direction within the domain. (b) Full computational domain and the mesh used in the outer region. The mesh consists of $281 \times 134 \times 113$ trilinear hexahedral elements in the streamwise, wall-normal, and circumferential directions, respectively. The grid is refined near the junction of the hemispherical nose and the cylindrical body to accurately capture cavitation inception and the complex interactions between the vapor cavity and the boundary layer. The mesh remains fixed throughout the simulation.}
    \label{fig:FlowOverHemisphere_mesh}
\end{figure}

We further consider an engineering-relevant application: partially cavitating flow over a three-dimensional cylindrical body with a hemispherical nose. We quantitatively compare the results obtained using {\MethodAbr} against experimental data \cite{McNown1948-rj}. The inner computational domain consists of a hemisphere with diameter $D = 2.54$ cm, attached to a congruent cylindrical body of length $50D$. The outer boundary of the domain is formed by a concentric hemispherical cap and cylindrical extension of diameter $50D$; see \cref{fig:FlowOverHemisphere_mesh}. We set the temperature $T = 295$ K and $F = 5\times 10^6$. To replicate the experimental setup, we impose a freestream pressure of $p_\infty = 22.6$ kPa and a freestream velocity $\bm{u}_\infty = [10, 0, 0]$ m/s. This yields a cavitation number
\begin{equation}
    \sigma = \frac{p_\infty - p^s(T)}{\frac{1}{2}\rho_\infty |\bm{u}|_\infty^2}
\end{equation}
equal to $0.4$. The Reynolds number based on the diameter $D$ is approximately $2.5 \times 10^5$, indicating a highly turbulent flow regime. At the interior boundary, we impose $\bm{u} = \bm{0}$ and $\nabla \rho\cdot\bm{n} = 0$, corresponding to a no-slip velocity condition and a $90^\circ$ contact angle between the wall and the liquid-vapor interface, respectively \cite{Hu2022-se}. At the outer boundary, we impose freestream velocity along the hemispherical cap and freestream pressure along the cylindrical extension. To minimize wave reflections from the outer domain, we added acoustically absorbing sponge layers of thickness $10D$ \cite{Colonius2004-rv, Zhou2010-ob}. Within these layers, source terms are introduced into the governing equations to damp both $\rho$ and $\bm{u}$ toward their freestream values. The initial condition is obtained from an incompressible flow simulation with constant density $\rho_\infty$.

\begin{figure}[htbp]
    \centering
    \includegraphics[keepaspectratio=true,width=\linewidth]{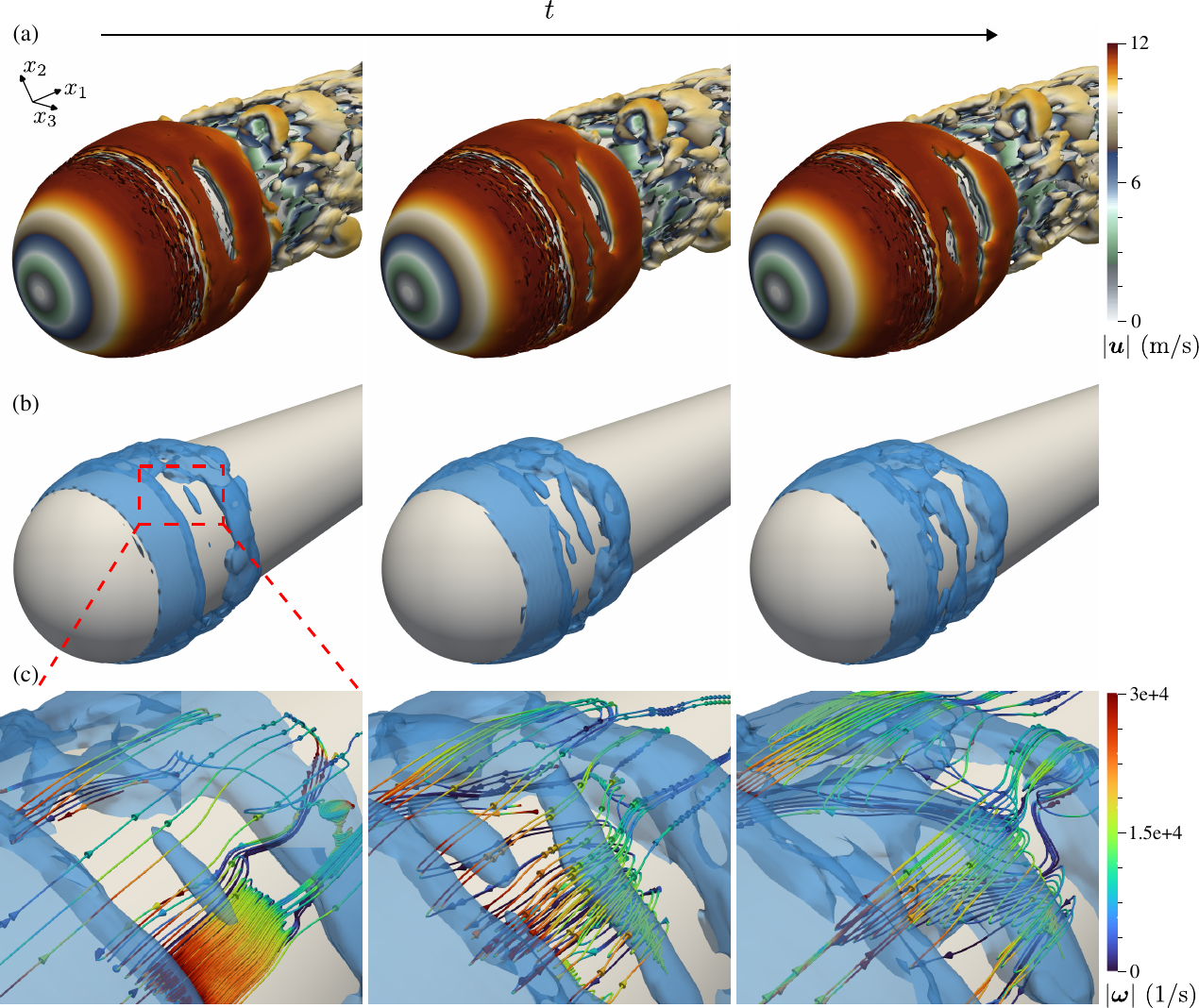}
    \caption{Instantaneous isocontours of (a) the Q-criterion at a value of 1550 s$^{-2}$, colored by velocity magnitude; (b) void fraction at $\alpha = 0.05$; and (c) a zoomed-in view of the void fraction, overlaid with velocity streamlines colored by vorticity magnitude. The flow is highly unsteady and involves periodic interactions between the vapor cavity and turbulent vortex cores.}
    \label{fig:FlowOverHemisphere_Inst}
\end{figure}

In \cref{fig:FlowOverHemisphere_Inst}(a), we show three snapshots of the isosurface of the Q-criterion at a value of 1550 s$^{-2}$, colored by velocity magnitude. While the flow accelerates smoothly over the hemispherical nose, cavitation induces the formation of vortex cores spanning multiple length scales. These vortices stretch and rotate as they move downstream, giving rise to complex, turbulent flow structures. \Cref{fig:FlowOverHemisphere_Inst}(b) displays instantaneous isocontours of the void fraction, defined as $\alpha = (\rho_l - \rho) / (\rho_l - \rho_v)$, at a value of 0.05. While a primary vapor sheet cavity remains attached to the bluff body, secondary cloud cavities are intermittently shedded near the trailing edge, exhibiting highly unsteady features. To further investigate the formation of cloud cavities, \cref{fig:FlowOverHemisphere_Inst}(c) presents a zoomed-in view of the instantaneous void fraction isocontour, overlaid with velocity streamlines colored by the vorticity magnitude $|\bm{\omega}|$. As small vapor pockets detach from the main cavity, they begin to interact with surrounding vortex structures. This interaction induces both rotation and stretching of the vapor pockets. As they grow, many of these small pockets merge into a larger cloud cavity with strong three-dimensional structures. The large cloud cavity is subsequently transported downstream by the vortex motion. Upon entering a higher-pressure region, it collapses violently, emitting strong shock waves that propagate upstream. These shock waves promote the shedding of additional small vapor pockets from the trailing edge of the primary sheet cavity, initiating a new cavitation cycle.

\begin{figure}[htbp]
    \centering
    \includegraphics[keepaspectratio=true,width=\linewidth]{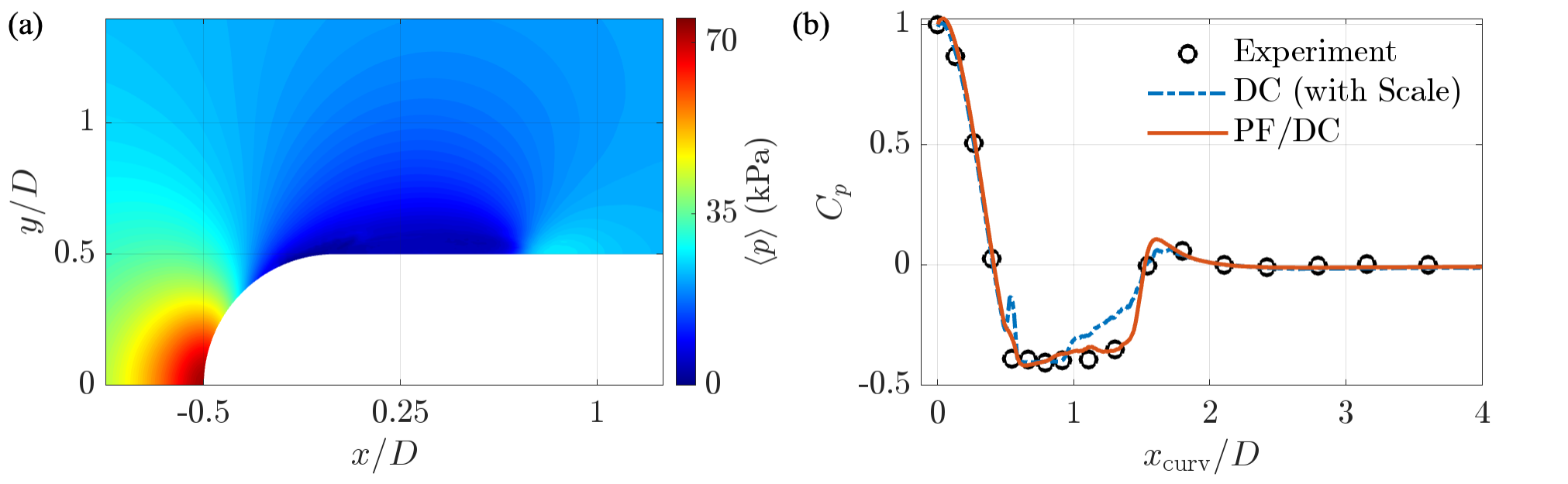}
    \caption{(a) Time- and circumferentially averaged pressure contours predicted by \cref{eqn:DVS_Formulation} using {\MethodAbr}. (b) Evolution of the pressure coefficient along the wall of the bluff body. Experimental data are taken from \cite{McNown1948-rj}, and the DC (with scale) results are taken from \cite{Hu2025-by}.}
    \label{fig:FlowOverHemisphere_Pressure}
\end{figure}

In \cref{fig:FlowOverHemisphere_Pressure}(a), we present the contour of time- and circumferentially averaged pressure, $\langle p \rangle$, around the cavitating region. As the fluid accelerates along the curved geometry, the pressure initially decreases. Once it falls below the saturation pressure, cavitation occurs and the pressure remains nearly constant at the saturation value over an extended region. At the end of the cavitating zone, the pressure rises sharply, forming a localized region with pressure exceeding the freestream value. Farther downstream, the pressure gradually relaxes back to the freestream condition. In \cref{fig:FlowOverHemisphere_Pressure}(b), we plot the wall distribution of the pressure coefficient, defined as $C_p = 2(p - p_\infty)/(\rho_\infty u_\infty^2)$, along the bluff body. We compare our results with experimental data from \cite{McNown1948-rj} and with numerical results obtained using the DC (with scale) operator \cite{Hu2025-by}. The {\MethodAbr} solution shows excellent agreement with the experimental measurements. In contrast, the DC (with scale) operator exhibits unphysical oscillations near $x_{\rm curv}/D = 0.5$ and predicts an earlier pressure recovery in the region $x_{\rm curv}/D \in [1, 1.5]$. Notably, the mesh used in this study contains only 76.8\% of the elements used in \cite{Hu2025-by}, yet the {\MethodAbr} operator still produces more accurate results compared to the classical DC formulation.

\begin{figure}[htbp]
    \centering
    \includegraphics[keepaspectratio=true,width=\linewidth]{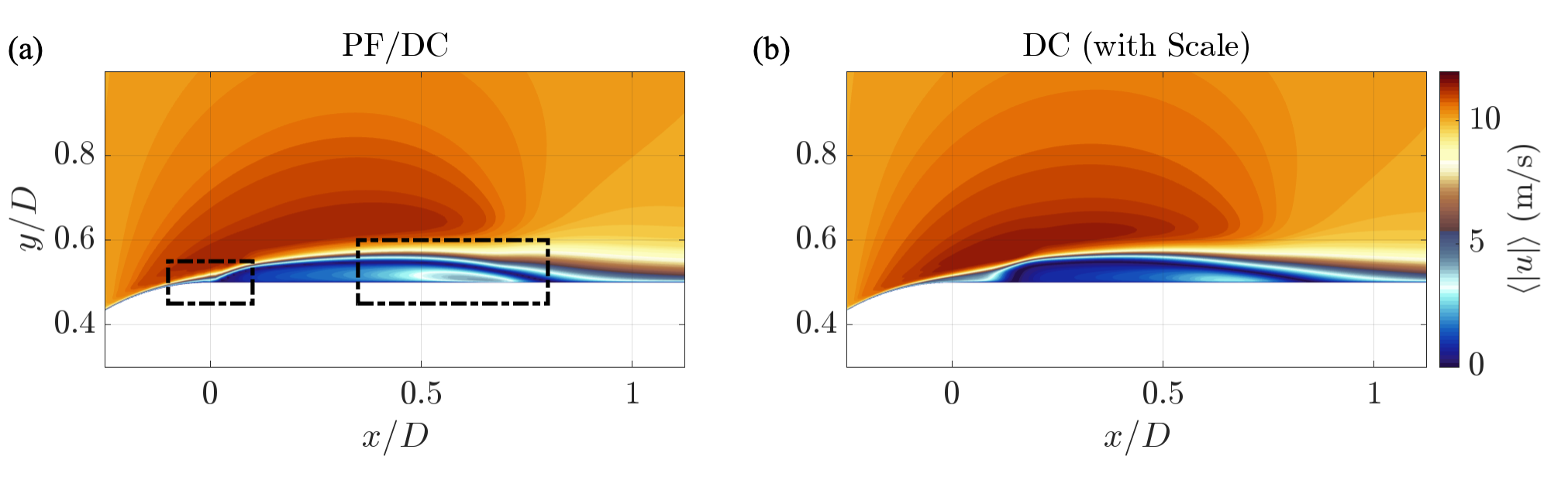}
    \caption{Time- and circumferentially averaged velocity magnitude contours predicted by \cref{eqn:DVS_Formulation} using (a) {\MethodAbr} and (b) the DC (with scale) operator, with (b) reproduced from \cite{Hu2025-by}. The black dashed boxes at $x/D = 0$ and $x/D = 0.5$ in (a) highlight that {\MethodAbr} predicts a much sharper boundary layer separation and a stronger jet upon flow reattachment to the wall, compared to the DC (with scale) operator.}
    \label{fig:FlowOverHemisphere_Velocity}
\end{figure}

\begin{figure}[htbp]
    \centering
    \includegraphics[keepaspectratio=true,width=\linewidth]{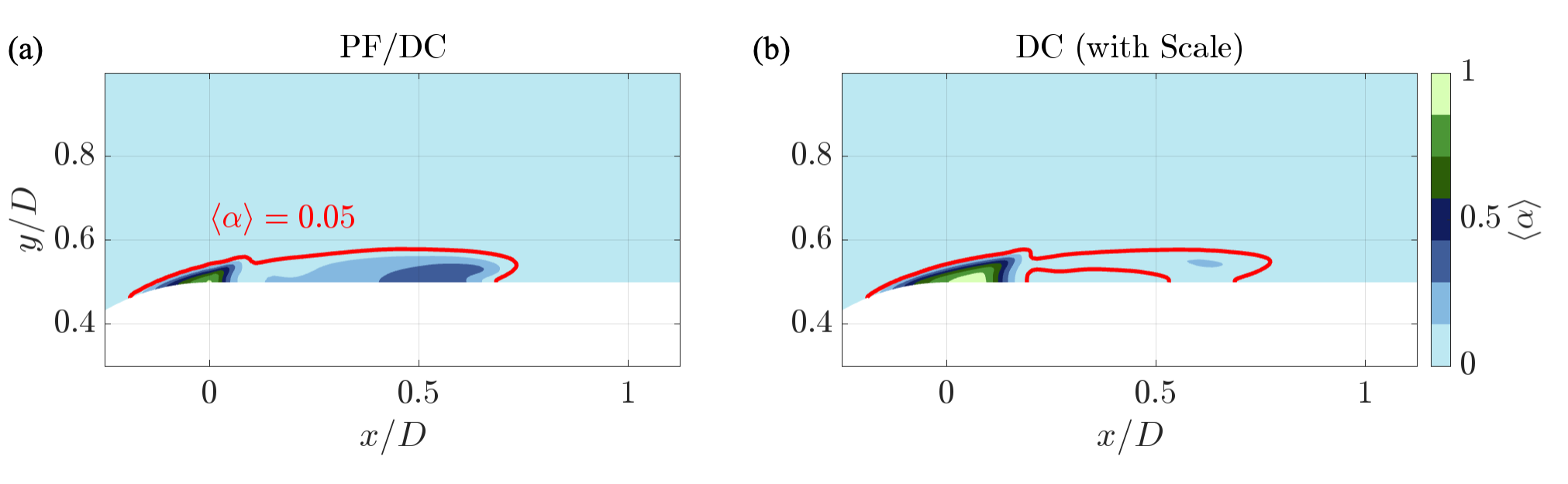}
    \caption{(a) Time- and circumferentially averaged void fraction contours predicted by \cref{eqn:DVS_Formulation} using (a) {\MethodAbr} and (b) DC (with scale) operator, with (b) reproduced from \cite{Hu2025-by}. The cavity shapes corresponding to the contour value $\langle \alpha \rangle = 0.05$ are marked in red.}
    \label{fig:FlowOverHemisphere_VoidFrac}
\end{figure}

Differences in pressure distribution can significantly influence the overall flow behavior. In \cref{fig:FlowOverHemisphere_Velocity}(a) and (b), we present the time- and circumferentially averaged velocity magnitude, $\langle |\bm{u}| \rangle$, obtained using {\MethodAbr} and the DC (with scale) operator, respectively. Near $x/D = 0$, {\MethodAbr} predicts a much sharper boundary layer separation compared to the DC (with scale) operator. While both methods yield a similar reattachment location near $x/D = 1$, {\MethodAbr} predicts a much stronger jet in the region $x/D \in [0.4, 0.75]$. We hypothesize that this jet leads to the rapid pressure rise at the end of the cavitating region, and is essential for accurately capturing the wall pressure distribution. In \cref{fig:FlowOverHemisphere_VoidFrac}(a) and (b), we show the time- and circumferentially averaged void fraction contours computed using {\MethodAbr} and the DC (with scale) operator, respectively. The cavity shapes corresponding to the contour value $\langle \alpha \rangle = 0.05$ are marked in red. Due to the premature pressure rise, the DC (with scale) operator predicts liquid pockets embedded within the cavity region. In contrast, {\MethodAbr} produces a continuous cavity outline from inception to termination.

\section{Conclusions and future work}\label{sec:concl}

In this paper, we present a new discontinuity-capturing operator to numerically solve flows with phase transformation. We begin by introducing the isothermal Navier–Stokes–Korteweg (NSK) equations, followed by a discussion of the choice of equation of state (EoS) and the associated interfacial dynamics. We show that the NSK equations admit the existence of an interfacial region where $\frac{dp}{d\rho} < 0$ and satisfy a free energy dissipation law for any arbitrary process. This property enables the model to capture both the coexistence of liquid and vapor phases and the non-equilibrium phase transition process without relying on any empirical mass transfer function.

We then discuss the necessity of a discontinuity-capturing (DC) operator in the numerical solution of compressible flows. To motivate the development of a new DC operator tailored for the NSK equations, we first review the classical design. Despite its success in gas dynamics, we show that the classical DC operator (1) introduces excessive dissipation in the bulk liquid and (2) violates the free energy dissipation law in the interfacial region when applied to solve NSK equations. We next examine recent modifications proposed in the literature that effectively mitigate the first issue but still fail to preserve the free energy dissipation law in the presence of liquid-vapor interfaces. Finally, we introduce the {\Methodfull} ({\MethodAbr}) operator, which integrates the concept of a nonlocal chemical potential from phase-field theory with traditional discontinuity capturing techniques for compressible flow. We demonstrate that {\MethodAbr} vanishes in the bulk liquid when the flow is nearly incompressible and satisfies the free energy dissipation law in the interfacial region.

We then apply the classical DC, its recent modifications, and {\MethodAbr} operators within the direct van der Waals simulation (DVS) framework. To stabilize numerical solutions involving mixed hyperbolic–dispersive wave structures, we also incorporate a dispersive-SUPG term into the complete numerical formulation. We demonstrate that only {\MethodAbr} retains the optimal rate of convergence in both the bulk liquid and interfacial regions. Moreover, {\MethodAbr} yields stable numerical solutions even when the liquid–vapor interface is severely under-resolved, whereas the classical DC operator generates unphysical wave structures and introduces excessive numerical dissipation. Finally, we apply the proposed algorithm to simulate cavitating flow over a 3D bluff body and show that the numerical results are in excellent agreement with experimental data.

In this paper, we focus on the isothermal case. While this assumption is reasonable for the interfacial dynamics and cavitation inception problems considered here, thermal effects become significant, particularly for phase transformations driven by temperature variations. To broaden the range of applications, we plan to extend both {\MethodAbr} operator and the DVS framework to thermally coupled phase-transforming flows. Additionally, we aim to generalize the DVS framework to binary mixtures, where both components can undergo phase transformation.

\section*{Declaration of competing interest}
The authors declare that they have no competing interests.

\section*{Acknowledgements}
This work is funded by the U.S. Department of Defense (award No. FA9550-20-1-0165 and N000142512096), PO Dr. Yin Lu (Julie) Young. This work uses the Bridges-2 system at the Pittsburgh Supercomputing Center (PSC) through allocation no. MCH220014 from the Advanced
Cyberinfrastructure Coordination Ecosystem: Services and Support (ACCESS) program, which is supported by the National Science Foundation, grant nos. 2138259, 2138286, 2138307,
2137603, and 2138296.

\section*{Data availability}
Data will be made available upon reasonable request.

\bibliographystyle{elsarticle-num}
\bibliography{paperpile.bib}

\newpage
\appendix
\setcounter{table}{0}
\renewcommand{\thetable}{A\arabic{table}}
\numberwithin{equation}{section}
\renewcommand{\thefigure}{A\arabic{figure}}
\setcounter{figure}{0}

\begin{appendices}
\section{Alternative design of {\Methodfull} operator}\label{apx:AlternativeDesign}

While in \cref{sec:PFDC} we introduce the {\MethodAbr} operator and show it satisfies the free energy dissipation law in the interfacial region, \cref{eqn:PFDC} is not the only possible design that exhibits this property. Still employing the concept of a nonlocal chemical potential, one of the most intuitive alternative designs is
\begin{equation}
    \bm{{F}}^{\rm DC}_i = \eta \kappa_{\rm DC}^{u} \begin{bmatrix} 
        0 \\ 
        \rho \bm{D}_{1i} \\ 
        \rho \bm{D}_{2i} \\ 
        \rho \bm{D}_{3i} 
    \end{bmatrix} + 
    \eta \kappa_{\rm DC}^{\mu} \begin{bmatrix} 
        \rho \mu_{{\rm nl},i} \\ 
        \rho u_1 \mu_{{\rm nl},i} \\ 
        \rho u_2 \mu_{{\rm nl},i} \\ 
        \rho u_3 \mu_{{\rm nl},i}
    \end{bmatrix},
    \label{eqn:PFDC_alternative}
\end{equation}
where the scaling coefficient $\eta$, artificial dissipation coefficients $\kappa_{\rm DC}^{u}$ and $\kappa_{\rm DC}^{\mu}$ are computed using \cref{eqn:DC_Scale,eqn:kappa_DC_u,eqn:kappa_DC_mu}, respectively. The corresponding free energy dissipation equation for \cref{eqn:DVS_CompactForm_Mod} with \cref{eqn:PFDC_alternative} is given by
\begin{equation}
    \frac{d}{dt} \mathcal{E}(\rho, \rho \bm{u}) = - \int_\Omega \nabla \bm{u} : \overline{\mu}(\rho) \bm{D} {\rm d}\Omega - \int_\Omega \nabla \bm{u} : \eta \rho \kappa_{\rm DC}^{u} \bm{D} {\rm d}\Omega - \int_\Omega\eta \rho \kappa_{\rm DC}^{\mu} |\nabla\mu_{\rm nl}|^2 {\rm d}\Omega \leq 0.
    \label{eqn:EnergyDissipation_PFDC_alternative}
\end{equation}
We can show that this alternative design ensures free energy stability in both the interfacial and bulk fluid regions for any arbitrary process. In addition, when the interface reaches equilibrium and $\nabla \mu_{\rm nl}$ vanishes, \cref{eqn:PFDC_alternative} eliminates any artificial dissipation induced by the density gradient, potentially leading to a more accurate interface profile. Finally, the alternative design avoids the discontinuous switching of dissipation flux between the bulk fluid and the interfacial region that occurs in \cref{eqn:PFDC}.

Despite the desirable properties of \cref{eqn:PFDC_alternative}, we find it to be much less robust compared to \cref{eqn:PFDC}, especially when the liquid-vapor interface is severely underresolved by the underlying mesh. To demonstrate this difference, we studied the advection of a vapor bubble with a near-equilibrium interface in a liquid jet, following the numerical procedures outlined in \cref{sec:num_procedure}. Here, we replace the modified GERG-2008 EoS with a double-well free energy potential with non-matching densities,
\begin{equation}
    f = \frac{\lambda}{L_{\rm inter}^2}\frac{(\rho - \rho_v^s)^2 (\rho - \rho_l^s)^2 }{\rho(\rho_v^s - \rho_l^s)^2},
    \label{eqn:double_well}
\end{equation}
where $L_{\rm inter}$ is the interface thickness. The one-dimensional equilibrium interface profile corresponding to \cref{eqn:double_well} can be analytically described by a hyperbolic tangent function, which helps minimize instabilities induced by an inaccurate initial interface. We set $L_{\rm inter} = 10\,\mathrm{nm}$, $\rho_v^s = 10\,\mathrm{kg/m}^3$, and $\rho_l^s = 1{,}000\,\mathrm{kg/m}^3$. The numerical study is performed on a two-dimensional domain of size $2\,\mu\mathrm{m} \times 1\,\mu\mathrm{m}$ with periodic boundary conditions in all directions. The initial density field is specified as
\begin{equation}
    \rho_0 \left(\bm{x},0\right) = \frac{\rho_v^s + \rho_l^s}{2} + \frac{\rho_v^s - \rho_l^s}{2}\tanh\left( \frac{ \sqrt{2} \left(\left|\bm{x} - \bm{x}_c\right| - R\right) }{ 2 L_{\rm inter} } \right),
\end{equation}
where $R = 0.25\,\mu\mathrm{m}$ is the initial vapor bubble radius and $\bm{x}_c = \left( 0.5, 0.5 \right)$ $\mu$m denotes the bubble center. We impose a uniform initial velocity field $\bm{u} \left( \bm{x}, 0 \right) = \left( 261.8, 0.0 \right)$ m/s, mimicking a high-speed liquid jet travelling horizontally across the domain.

\begin{figure}[htbp]
    \centering
    \includegraphics[keepaspectratio=true,width=\linewidth]{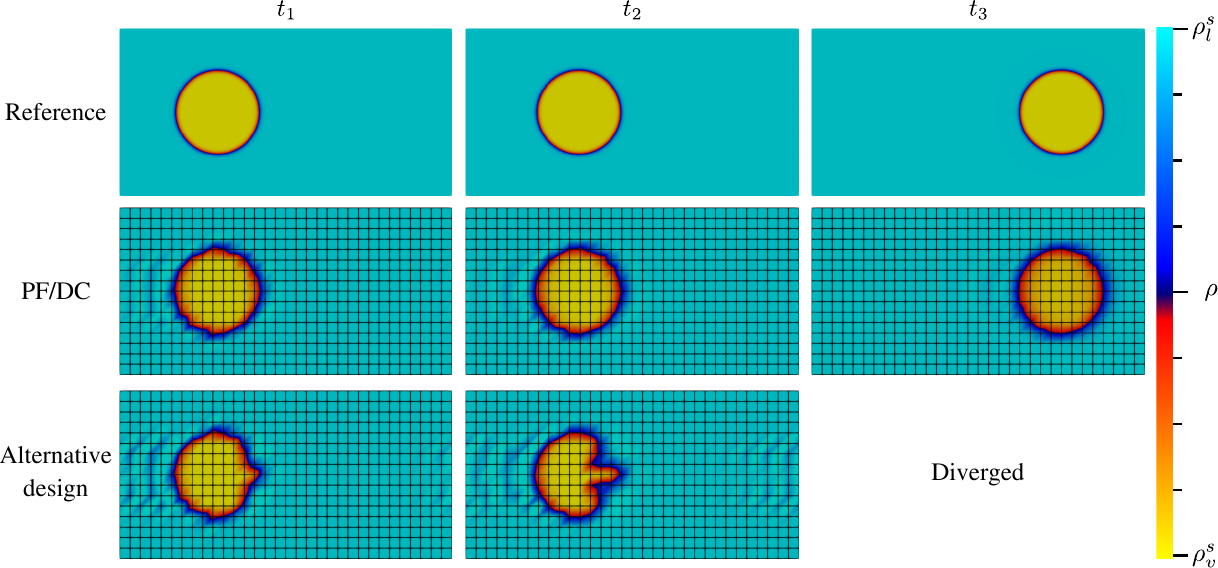}
    \caption{Time evolution of the density field for the reference solution (top row), the solution obtained with {\MethodAbr} (middle row), and the solution obtained with the alternative design (bottom row). The black lines overlaid on the contours in the middle and bottom rows indicate the underlying mesh.}
    \label{fig:BubbleTransport_Density}
\end{figure}

\begin{figure}[htbp]
    \centering
    \includegraphics[keepaspectratio=true,width=\linewidth]{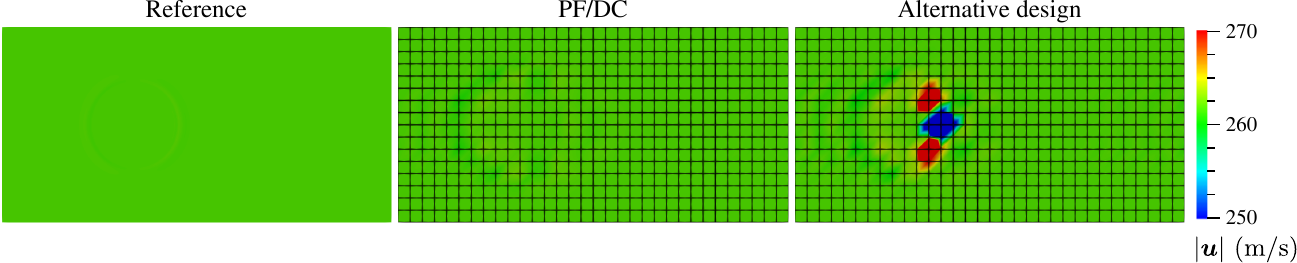}
    \caption{Velocity field contours at time $t_2$ for the reference solution (left), the solution computed with {\MethodAbr} (middle), and the solution computed with the alternative design (right). The black lines overlaid on the contours in the middle and right figures indicate the underlying mesh.}
    \label{fig:BubbleTransport_Velocity}
\end{figure}

The reference solution is computed by solving \cref{eqn:DVS_Formulation} without any DC operator on a fine mesh of $512 \times 256$ bilinear elements, and the time evolution of the density field is shown in \cref{fig:BubbleTransport_Density}. Since the initial density profile is close to equilibrium, the vapor bubble maintains its original shape and translates uniformly as the simulation progresses. For comparison, we also solve \cref{eqn:DVS_Formulation} with the {\MethodAbr} operator \cref{eqn:PFDC} and the alternative design \cref{eqn:PFDC_alternative} on a coarse mesh of $32 \times 16$ bilinear elements. On this coarse mesh, the interface thickness is smaller than the width of a single element, posing significant challenges on the numerical algorithm. As shown in \cref{fig:BubbleTransport_Density}, despite some initial oscillations, {\MethodAbr} preserves a nearly circular bubble shape and accurately captures its movement with the high-speed liquid jet. In contrast, the alternative design exhibits local spikes at the interface at time $t_1$. As the simulation progresses, these spikes grow, resulting in a highly irregular bubble shape at $t_2$ and ultimately leading to divergence at $t_3$. To illustrate the cause of divergence, we plot the velocity field contours at $t_2$ in \cref{fig:BubbleTransport_Velocity}. While {\MethodAbr} produces a nearly uniform velocity field, closely matching the reference solution, the alternative design generates strong oscillations, which subsequently lead to solver failure. These results provide further evidence supporting the design of {\MethodAbr} in \cref{eqn:PFDC} and demonstrate the robustness of the proposed method.

\newpage

\end{appendices}

\end{document}